\documentclass[twocolumn,showpacs,preprintnumbers,amsmath,amssymb,superscriptaddress]{revtex4-1}

\usepackage{graphicx}
\usepackage{dcolumn}
\usepackage{bm}
\usepackage[version=3]{mhchem} 
\usepackage{amsmath}
\usepackage{esvect}
\usepackage{import}
\usepackage[draft]{changes}

\usepackage{array}
\usepackage{graphicx}
\usepackage{graphicx}
\usepackage{dcolumn}
\usepackage{bm}
\usepackage{color}
\usepackage{pifont}
\usepackage{natbib}
\usepackage{hyperref}
\hypersetup{
colorlinks = true,
urlcolor   = blue,
linkcolor  = blue,
citecolor  = blue
}
\usepackage{nicefrac}
\usepackage{changes}
\setremarkmarkup{(#2)}

\newcommand{\knto}{\ce{K2Ni2TeO6}}
\newcommand{\cc}{\textit{c}}
\newcommand{\x}{\textit{x}}

\newcommand{\ac}{\textit{a}}

\newcommand{\Vc}{\textit{V}}

\newcommand{\doc}{$^{\circ}$C}

\newcommand{\TN}{$T_{\rm N}$}
\newcommand{\musr}{$\mu^+$SR}

\begin{document}

\preprint{}
\title{Magnetism and Ion Diffusion in Honeycomb Layered Oxide \knto:\\ First Time Study by Muon Spin Rotation \& Neutron Scattering}


\author{N.~Matsubara}
 \email{namim@kth.se}
\author{E.~Nocerino}
\author{O.~K.~Forslund}
\author{A.~Zubayer}
\affiliation{Department of Applied Physics, KTH Royal Institute of Technology, SE-10691 Stockholm, Sweden}
\author{K.~Papadopoulos}
\affiliation{Department of Physics, Chalmers University of Technology, SE-41296 G{\"o}thenburg, Sweden}
\author{D.~Andreica}
\affiliation{Faculty of Physics, Babeş-Bolyai University, 400084 Cluj-Napoca, Romania}
\author{J.~Sugiyama}
\affiliation{Neutron Science and Technology Center, Comprehensive Research Organization for Science and Society (CROSS), Tokai, Ibaraki 319-1106, Japan}
\author{R.~Palm}
\affiliation{Department of Applied Physics, KTH Royal Institute of Technology, SE-10691 Stockholm, Sweden}
\author{Z.~Guguchia}
\affiliation{Laboratory for Muon Spin Spectroscopy, Paul Scherrer Institute, CH-5232 Villigen PSI, Switzerland}
\author{S.P.~Cottrell}
\affiliation{ISIS Muon Facility, Rutherford Appleton Laboratory, Didcot, Oxfordshire, OX11 0QX, UK}
\author{T.~Kamiyama}
\affiliation{Institute of Materials Structure Science, High Energy Accelerator Research Organization, 203-1 Shirakata, Tokai, Ibaraki 319-1107, Japan}
\author{T.~Saito}
\affiliation{Institute of Materials Structure Science, High Energy Accelerator Research Organization, 203-1 Shirakata, Tokai, Ibaraki 319-1107, Japan}
\author{A.~Kalaboukhov}
\affiliation{Microtechnology and Nanoscience, Chalmers University of Technology, SE-41296 G{\"o}thenburg, Sweden}
\author{Y.~Sassa}
\affiliation{Department of Physics, Chalmers University of Technology, SE-41296 G{\"o}thenburg, Sweden}
\author{T.~Masese}
\affiliation{Department of Energy and Environment, Research Institute of Electrochemical Energy (RIECEN), National Institute of Advanced Industrial Science and Technology (AIST), Ikeda, Osaka 563-8577, Japan}
\affiliation{AIST-Kyoto University Chemical Energy Materials Open Innovation Laboratory (ChEM-OIL), National Institute of Advanced Industrial Science and Technology (AIST), Sakyo-ku, Kyoto 606-8501, Japan}
\author{M.~M{\aa}nsson}
 \email{condmat@kth.se}
\affiliation{Department of Applied Physics, KTH Royal Institute of Technology, SE-10691 Stockholm, Sweden}

\date{\today}

\begin{abstract}
In the quest of finding novel and efficient batteries, a great interest has raised in K-based honeycomb layer oxide materials both for their fundamental properties and potential applications. A key issue in the realization of efficient batteries based on such compounds, is to understand the K-ion diffusion mechanism. However, investigation of potassium-ion (K$^+$) dynamics in materials using magneto-spin properties has so far been challenging, due to its inherently weak nuclear magnetic moment, in contrast to other alkali ions such as lithium and sodium. Spin-polarised muons, having a high gyromagnetic ratio, make the muon spin rotation and relaxation (\musr) technique ideal for probing ions dynamics in weak magneto-spin moment materials. Here we report the magnetic properties and K$^+$ dynamics in honeycomb layered oxide material of the \knto~using \musr~measurements. Our low-temperature \musr~results together with, with complementary magnetic susceptibility, find an antiferromagnetic transition at 26~K. Further \musr~studies performed at higher temperatures reveal that potassium ions (K$^+$) become mobile above 250~K and the activation energy for the diffusion process is $E_a= 121 (13)$~meV. This is the first time that K$^+$ dynamics in potassium-based battery materials has been measured using \musr. Finally our results also indicate an interesting possibility that K-ion self diffusion occurs predominantly at the surface of the powder particles. This opens future possibilities for improving ion diffusion and device performance using nano-structuring.

\end{abstract}

\keywords{layered honeycomb structure, potassium battery, \musr}
\maketitle
\newpage


\section{\label{sec:Intro}Introduction}

 \begin{figure*}[t]
   \begin{center}
     \includegraphics[width=150mm]{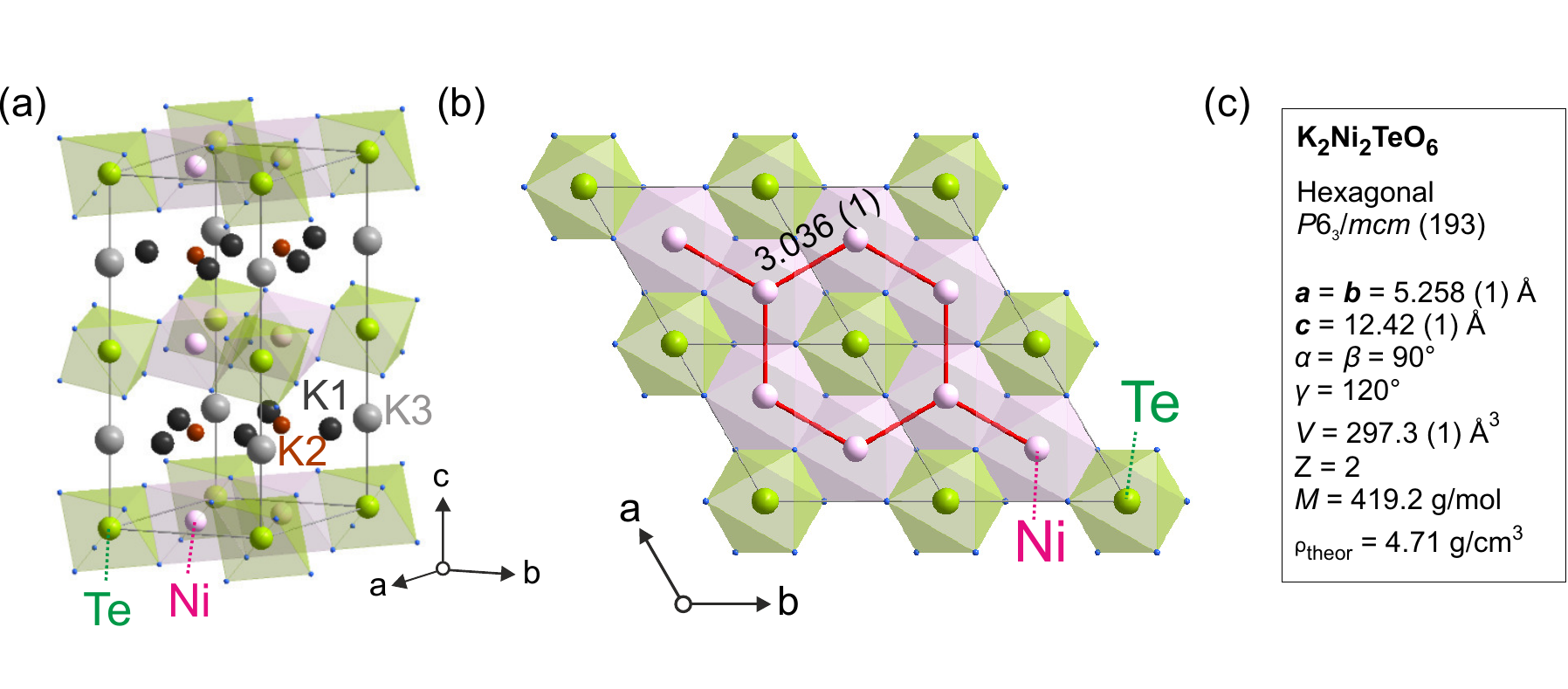}
   \end{center}
   \caption{(a,b) Honeycomb layered structure of \knto~showing the Ni atoms (pink), Te atoms (green), O atoms (blue) and K atoms occupying crystallographic sites (K1 (black), K2 (red), and K3 (gray)) that can move in a two-dimensional (2D) fashion, as reported in Ref.\cite{Masese2018}. Red line indicates Ni-Ni network (interatomic distance in \AA). (c) The crystal structure information of \knto~is extracted from room temperature-NPD data (Fig. \ref{NPDXRD}(a,b)) $Z$ is the number of formula units in the unit cell. }
   \label{structure}
 \end{figure*}

Layered oxides have attracted prime attention over the last decade owing to their intriguing physical and chemical properties across varied aspects to, phase transitions (e.g. antiferromagnetism, superconductivity, Kitaev magnet)\cite{Matsubara2019c, Lee2017, Karna2017, Kim2016a}, thermodynamics (e.g. fast ionic conductivity) \cite{Yang2017a} and unusual electromagnetic spin interactions  (multiferroics, high-voltage electrochemistry) \cite{Khanh2016, Chaudhary2018, Choi2019}. Layered oxides consisting of alkali atoms sandwiched between slabs with transition metal atoms (commonly referred to as layered transition metal oxides, TMOs), have been extensively investigated. This is especially true for TMOs adopting the chemical composition $AM$\ce{O2}, where $A$ denotes an alkali atom and $M$ is typically a transition metal atom. Such compounds have raised interests not only from a fundamental point of view but also for applications. For instance, \ce{Na_xCoO2} are widely known materials exhibiting rich of intriguing physical properties at low temperatures such as spin density waves \cite{Wooldridge2005}, superconductivity (hydrated compound \cite{Schaak2003, Takada2009}), metal-insulator transitions \cite{Sugiyama2004}, in addition to unique magnetic and charge ordering phases \cite{Hertz2008}.

In electrochemistry, \ce{Na_xCoO2} has also been investigated as a cathode material in Na-ion batteries, not only from environmental point of view but also for its fast sodium-ion diffusive capabilities \cite{Mansson2020, RamiReddy2015, Rai2014}. Despite this, another class of layered oxides has emerged to supersede \ce{Na_xCoO2}, such as \ce{Na2Ni2TeO6} (or equivalently as \ce{Na_{2/3}Ni_{2/3}Te_{1/3}O_2}) \cite{Gupta2013a, Berthelot2012} and \knto~(\ce{K2/3Ni2/3Te1/3O2}) \cite{Masese2018}, which show higher voltage (vs Na/Na$^+$ ; K/K$^+$) cation electrochemistry and better structural stability.

\knto~adopts essentially the same crystal structure as \ce{Na2Ni2TeO6},  
with a significant increase of the interslab distance owing the larger potassium atoms to reside in between the honeycomb slabs of Ni octahedra surrounding Te octahedra in a honeycomb fashion (Fig. \ref{structure}(a,b)). In addition, interesting magnetic properties are anticipated in \knto, arising from the regular honeycomb configuration of Ni atoms, which triggers the interplay between adjacent antiferromagnetic (AFM) interactions \cite{Karna2017, Berthelot2012}. Moreover, complex magnetic structures can be expected owing to the competition between the direct interactions of magnetic Ni atoms and exchange interactions through the non-magnetic atoms.

The large ionic radii of potassium cations, which leads to the increase in the interslab distance, influences not only the electric and spin interactions but also the K diffusion properties. 

In contrast to Na and Li, K has a weak nuclear magnetic moment that makes this interaction difficult to probe. This places muon spin rotation and relaxation (\musr) measurements at the frontier techniques for probing the aforementioned interactions, since the positive muon possesses a charge and a high gyromagnetic ratio. In particular, the positive muon is typically bound to the negatively charged oxygen atoms at a distance of 1~\AA, and interact with both nuclear and electronic moments in the matter. This means that the muon may couple even to the weak nuclear moment of K, given the high gyromagnetic ratio of the muon. Here we report first measurements of magnetic properties and K-ion dynamics in honeycomb layered \knto~oxide material using \musr. Room-temperature x-ray and neutron powder diffraction experiments confirm that the average crystal structure is in agreement with the reported one. \knto~exhibits an antiferromagnetic transition at 26~K and ZF-\musr~oscillation signal suggests commensurate spin ordering down to 2~K. \musr~studies performed on \knto~at high temperatures reveal that potassium ions (K$^+$) are dynamic above 250~K (with an activation energy of 121 (13)~meV extracted from the experimental data), revealing for the first time that K$^+$ dynamics can potentially be measured using \musr.

\section{\label{sec:results}Results and Discussion}
 \subsection{\label{sec:level2}  Room temperature diffraction} 
\begin{figure}[ht]
   \begin{center}
     \includegraphics[width=75mm]{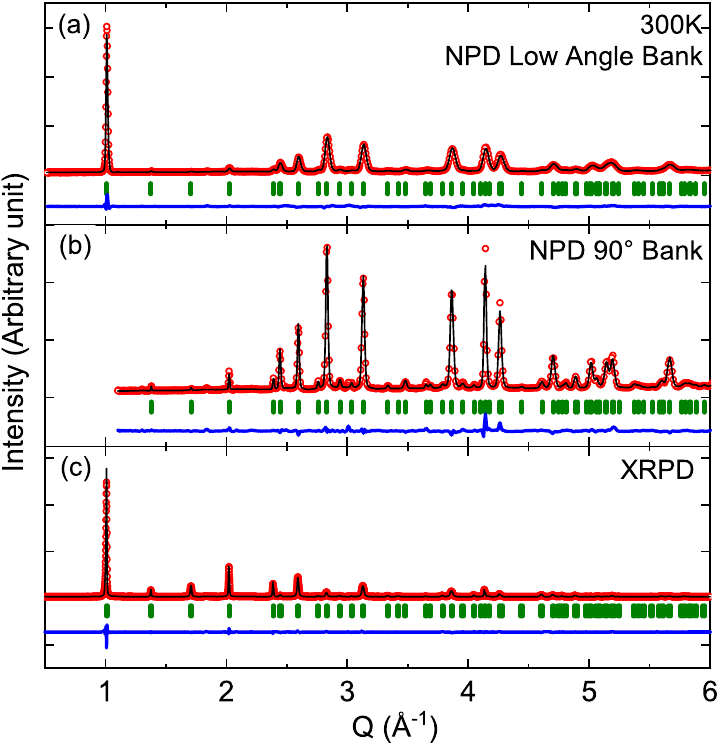}
   \end{center}
   \caption{Rietveld refinements of (a) NPD low angle bank, (b) 90$^{\circ}$ bank and (c) XRPD of \knto~at 300~K.}
   \label{NPDXRD}
 \end{figure}

The crystal structure of \knto~at room temperature (300~K) was obtained by refinements of both x-ray powder diffraction (XRPD) and neutron powder diffraction (NPD) data. The structural parameters of \knto~started from the reported unit cell ($P6_3/mcm$ with $a$ = 5.26 \AA, $c$ = 12.47 \AA) and atomic coordinates \cite{Masese2018}. The Rietveld fits of high-resolution neutron powder diffraction patterns was challenging due to the significant broadening observed on [$h$, $k$, $l$ $\neq$ 0] peaks. Similar broadening profile was reported on \ce{Na2Ni2TeO6}, for which Karba \textit{et al.} introduced the anisotropy strain into consideration to the crystal structure refinement process \cite{Karna2017}. In our report here, we also used the anisotropic strains based on a spherical harmonics modelling of the Bragg peak broadening using the Fullprof suite in order to fit both XRPD and NPD data. As Karba \textit{et al.} pointed out, this strong broadening is probably originated from both the anisotropic displacement of oxygen atoms under thermal fluctuation and the potential gradient of Na/K distribution due to the weaker interlayer interaction in this type of structure. The average crystal structure model provides fitting to both the XRPD and NPD data, as shown in both Fig. \ref{NPDXRD} as well as the detailed refinement data and the corresponding structure obtained by NPD are supplied in Fig. \ref{structure} and the supplementary materials (Table. \textcolor{blue}{S1 and S2}). The average  structure is consistent with the reported one \cite{Masese2018}.  Note that the accurate crystal structure determination is beyond the scope of this paper, thus the obtained average structure model of \knto~will be used for the estimation of the K-ion diffusion coefficient as detailed below in Section \ref{HT_MUSR}.


 \subsection{\label{sec:level2}  Magnetic susceptibility} 

\begin{figure}[ht]
   \begin{center}
     \includegraphics[width=75mm]{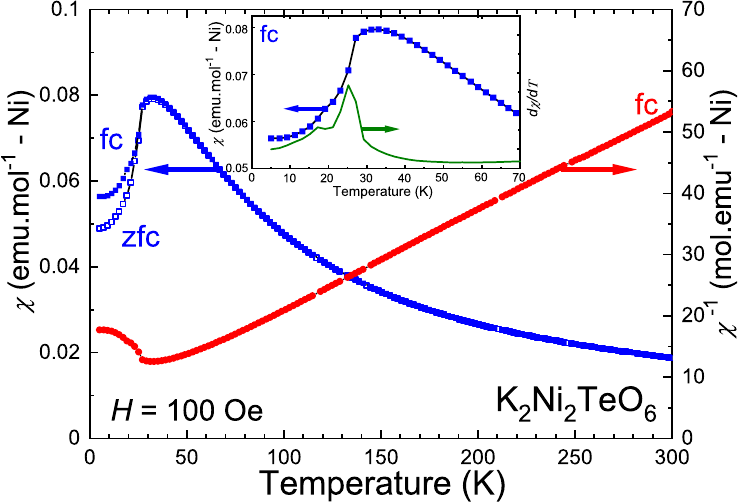}
   \end{center}
   \caption{Magnetic susceptibility ($\chi$ ($T$) and 1/$\chi$($T$)) curves of \knto~recorded in zero-field-cooled ($zfc$) and field-cooled ($fc$) modes under an applied magnetic field of 100 Oe, with the corresponding Curie-Weiss fitting as a dotted line. Inset shows the magnified image of the susceptibility plot and of the corresponding differential susceptibility  [$d\chi$/$dT$]($T$) curve indicating \TN~= 26 K. }
   \label{sus}
 \end{figure}

Fig. \ref{sus} shows the DC magnetic susceptibility of \knto~measured under a magnetic field of 100 Oe in the temperature range $T$ = 5 - 300~K. \knto~exhibits AFM behaviour with a maximum of the $\chi$ curve at around 33~K. The magnetic transition is also well evidenced in the differential susceptibility [$d \chi$/$dT$]($T$) curves, at the AFM N\'{e}el temperature $T_{\rm N}\approx26$~K in both $zfc$ and $fc$ modes (only $fc$ is shown in inset of Fig. \ref{sus}).

No significant divergence between $zfc$ and $fc$ magnetisation curves is observed. The susceptibility data (1/$\chi$) were fitted with a Curie-Weiss law (using data points above 80~K), yielding a Weiss temperature $\theta_{\rm CW}$ = -30.3 K. The negative Weiss temperature indicates AFM interactions, which could arise from the superexchange interactions between the nearest and the next-nearest neighbours of Ni. Further, an effective magnetic moment, $\mu_{\rm eff}$ = 2.53 $\mu_{\rm B}$/Ni was obtained, which is in excellent agreement with the theoretical spin only value for ${\rm Ni}^{2+}$ (2.83 $\mu_{\rm B}$).



 \subsection{\label{sec:level2}  Low-temperature wTF \musr~ measurements} 
 
 \begin{figure*}[ht]
   \begin{center}
     \includegraphics[width=\linewidth]{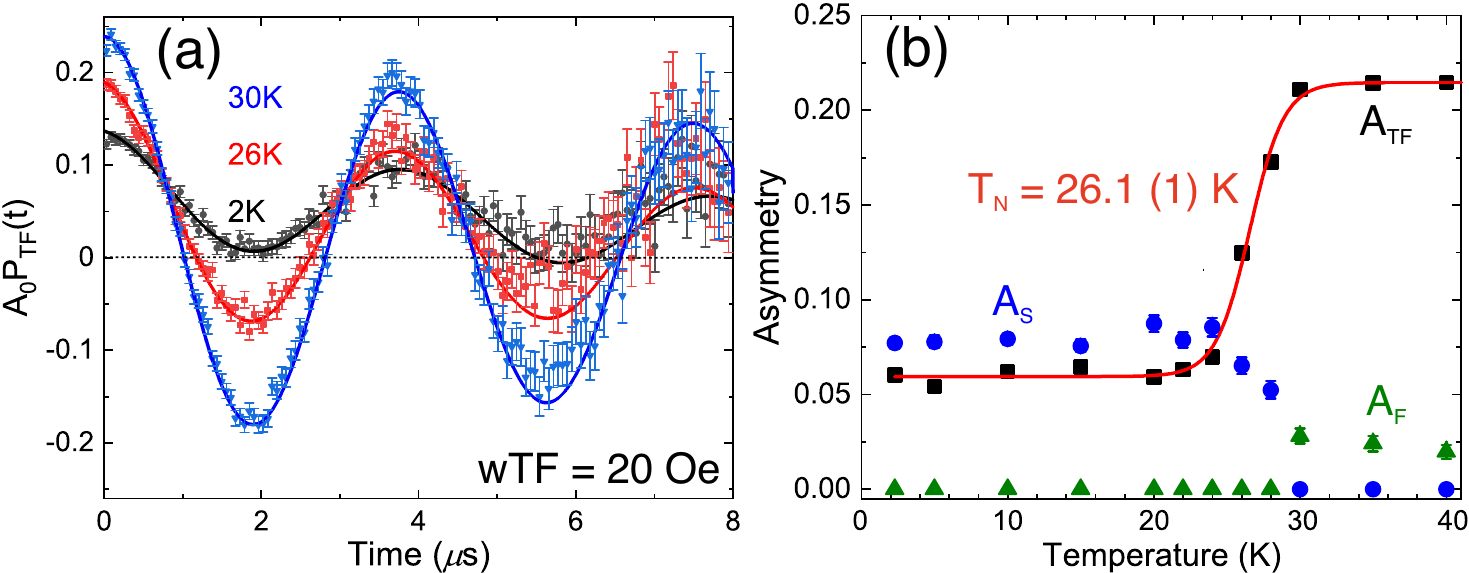}
   \end{center}
   \caption{(a) \musr~time spectra measured at temperatures of 2, 26 and 30~K under a weak-transverse field (wTF) of 20 Oe,  and the corresponding fits using Eq. \ref{1} (solid lines). For clarity, $A_0$ is the initial asymmetry and $P_{\rm TF}(t)$ is the muon spin polarisation function. (b) Asymmetry plots as functions of temperature, where $A_{\rm TF}$, $A_{\rm S}$ and $A_{\rm F}$ are the asymmetries of the related polarisation components. The sigmoid fit (red solid line) indicated the antiferromagnetic transition temperature (\TN) at 26.1 K.} 
   \label{wTF}
 \end{figure*}

Fig. \ref{wTF}(a) shows the wTF \musr-time spectra recorded with $H$ = 20~Oe for three selected temperatures. The spectra was obtained on GPS in PSI without sample cell. Here, wTF means that the field is perpendicular to the initial muon spin polarization and its magnitude is very small compared with the internal magnetic field ($H_{\rm int}$) generated by magnetic order and/or disorder. When the temperature decreases below 30~K, the oscillation amplitude due to the applied wTF rapidly decreases, indicating the appearance of additional internal magnetic fields (i.e. static magnetic order). Below 30~K, the wTF \musr~time spectrum was consequently fitted using a combination of an exponentially relaxing precessing component and a slow-exponentially relaxing non-oscillatory component. The first component comes from the muons stopped in paramagnetic phases, where the internal magnetic field is equivalent to wTF = 20~Oe.
The second component corresponds to the magnetically ordered phase, where $H_{\rm int}$ >> wTF. Above 30 K, the wTF spectrum was also fitted using a combination of an exponentially relaxing oscillating and a fast-exponentially relaxing non-oscillatory components, probably due to fast spin fluctuation above \TN. The resulting fit function for the wTF spectra in the wide temperature range across \TN~is as follows:

 \begin{eqnarray}
A_0 \, P_{\rm TF}(t) &=& A_{\rm TF}\cos(2\pi f_{\rm TF}t + \phi_{\rm TF})\times{}\exp{(-\lambda_{\rm TF} t)}
\cr
&+& A_{\rm S}\times{}\exp{(-\lambda_{\rm S} t)}
\cr
&+& A_{\rm F}\times{}\exp{(-\lambda_{\rm F} t)},
\label{1}
\end{eqnarray}

where $A_S$ = 0 at $T$ $\geq$ \TN, while $A_F$ = 0 at $T$ $\leq$ \TN.
$P_{\rm TF}(t)$ is the muon spin polarisation function, $A_0$ is the initial asymmetry,  $A_{\rm TF}$, $A_{\rm S}$ and $A_{\rm F}$ are the asymmetries of the related polarisation components, $2\pi f_{\rm TF}$ is the angular frequency of the Larmor precession under the applied wTF, $\lambda_{\rm TF}$, $\lambda_{\rm S}$ and $\lambda_{\rm F}$ are the exponential relaxation rates for the three components and $\phi$ is the initial phase of the processing signal. 

The magnetic transition temperature is obtained from the $A_{\rm TF}$($T$) curve, because $A_{\rm TF}$ corresponds to the paramagnetic (PM) fraction of the sample. Thus, a step-like change in the $A_{\rm TF}$($T$) curve around 27~K indicates a transition from a low-temperature magnetically ordered state to a high-temperature PM state. As shown in Fig. \ref{wTF} (b), $A_{\rm TF}$($T$) dependence has been fitted with a sigmoid function and the transition temperature is defined as the middle point of the fitting curve, 
i.e. \TN~= 26.1 (1) K, which is in excellent agreement with the \TN~determined by magnetisation measurement (Fig.~\ref{sus}).

Below 20~K down to 2~K, the oscillation from the externally applied field is still clearly observed (see black curve in Fig.~\ref{wTF}(a)), having a volume fraction of about 28 $\%$, corresponding to a PM phase even at $T$ = 2~K. This suggests the presence of a second paramagnetic phase at even low temperatures. The absence of any detectable major impurity phases from diffraction measurements implies that the crystal structure of the second phase is very similar to that of the predominant phase. This could be related to the broadening observed in high-resolution NPD data due to the distribution of atoms in the structure. Such scenario could lead to atomic and magnetic order/disorder transitions at low temperatures \cite{Schulze2008}. Further, studies using $^3$He or dilution crysostats will be needed to clarify this interesting issue. This is further supported by the fact that the K-ions are mobile at room temperature (see below).

 \subsection{\label{sec:level2} Low-temperature ZF \musr~ measurements} 

\begin{figure}[ht]
   \begin{center}
     \includegraphics[width=75mm]{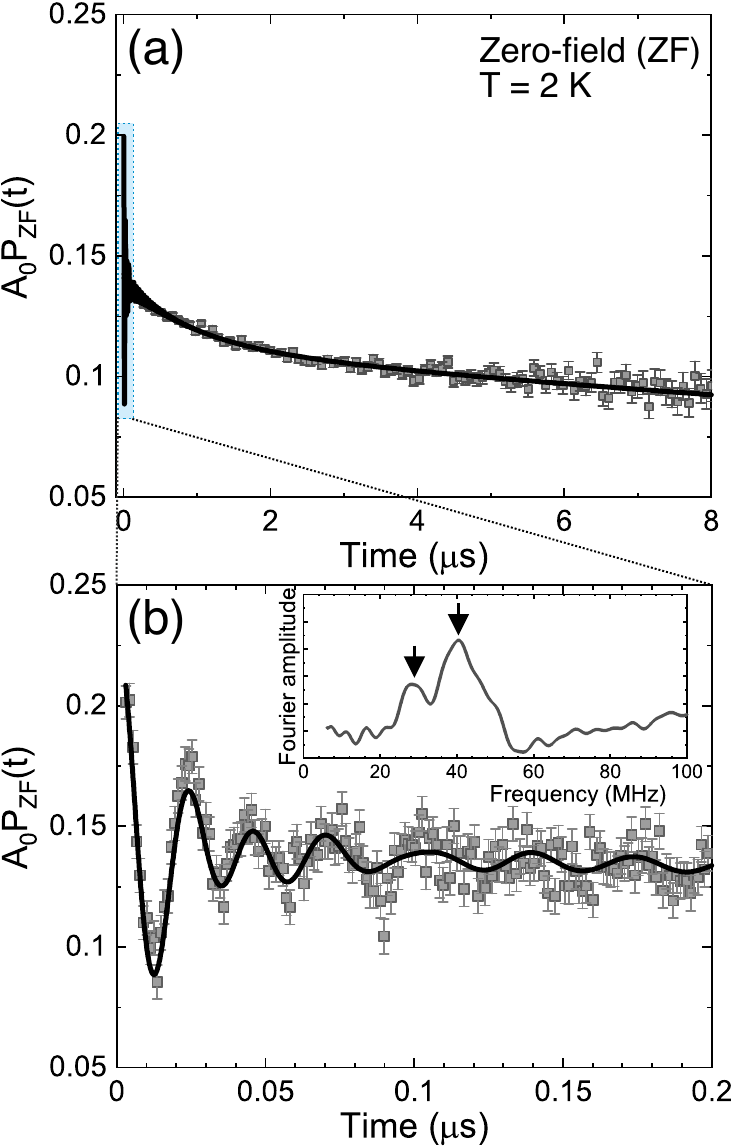}
   \end{center}
   \caption{The ZF-\musr~time spectrum recorded at a temperature of 2~K under zero-field (ZF): (a) in the long time domain up to 8~$\mu$s and (b) in an early time domain up to 0.2~$\mu$s. The inset of (b) shows a shorter time domain where the muon spin precession is evident the Fourier transformed frequency spectrum. Two frequencies ($f_{\rm AF1}$ = 29~MHz and $f_{\rm AF2}$ = 43~MHz) are pointed by black arrows. The Solid lines are represent the best fits of the data using Eq.~\ref{2}.}
   \label{ZF}
 \end{figure}

To further understand the magnetic nature of \knto, zero-field (ZF) \musr~ measurements were performed at temperatures between 2 and 40~K. As seen in Fig.~\ref{ZF}, the ZF-\musr~time spectra recorded at 2~K clearly shows the muon spin precession signal, which evidences the appearance of quasi-static magnetic order. 
Fourier transform of the ZF-\musr~ time spectrum (inset of Fig. \ref{ZF}(b)) reveals the presence of two distinct components namely: $f_{\rm AF1}$ = 29~MHz and $f_{\rm AF2}$ = 43~MHz, with an asymmetry ratio of 1 : 10 (as is shown in Fig. \ref{ZF_para}(d)).

In addition, there is a fast relaxing signal in the initial time spectra (see also Fig. \textcolor{blue}{S1}). Such behaviour may have several explanations; in particular this kind of signal might be due to delocalised muons or fast fluctuating moments arising from either the Ni ions or magnetic impurities. Thus, this ZF spectrum at 2 K was fitted by a combination of two exponentially relaxing cosine oscillations, which are originating from the magnetic order, one fast and one slow (for 1/3 powder average tail) exponentially relaxing non-oscillatory components and one exponentially relaxing non-oscillatory components due to the PM signal observed in the wTF measurement ($A_{\rm PM}$ fixed at 0.0728):  
\begin{eqnarray}
A_0 \, P_{\rm ZF}(t) &=&
A_{\rm AF1}\cos(2\pi f_{\rm AF1} t + {\phi_{\rm AF1}})\times{}\exp{(-\lambda_{\rm AF1} t)}
\cr
 &+& A_{\rm AF2}\cos(2\pi f_{\rm AF2} t + {\phi_{\rm AF2}})\times{}\exp{(-\lambda_{\rm AF2} t)}
\cr
 &+& A_{\rm F}\times{}\exp{(-\lambda_{\rm F} t)}
 \cr
 &+& A_{\rm tail}\times{}\exp{(-\lambda_{\rm tail} t)}
 \cr
 &+& A_{\rm PM}\times{}\exp{(-\lambda_{\rm PM} t)},
\label{2}
\end{eqnarray}
where $A_0$ is the initial asymmetry, $A_{\rm AF1}$, $A_{\rm AF2}$, $A_{\rm F}$, $A_{\rm tail}$ and $A_{\rm PM}$ are the asymmetries associated with each signals, $f_{\rm AF}{_i}$ is the frequency of the muon spin precession corresponding to the static internal AF field, $\phi_{\rm AF}{_i}$ is the initial phase of the oscillatory signal, $\lambda_{\rm AF}{_i}$, $\lambda_{\rm F}$, $\lambda_{\rm tail}$ and $\lambda_{\rm PM}$ are the exponential relaxation rates of each signal. $A_{\rm PM}$ was fixed at 0.0728, based on wTF measurements. As clearly shown in Fig. \ref{ZF}, the ZF-\musr~time spectrum is well fitted using Eq.~\ref{2} both in short ($t$ $\leq$ 0.2 $\mu$s) and long ($t$ $<$ 8 $\mu$s) time domain.

Both $\phi_{\rm AF1}$ and $\phi_{\rm AF2}$ show similar temperature trend from individual fitting (not shown), thus a common $\phi_{\rm AF}$ was finally used in the fitting, i.e. $\phi_{\rm AF}$ = $\phi_{\rm AF1}$ = $\phi_{\rm AF2}$. Both $A_{\rm AF1}$ and $A_{\rm AF2}$ were also found to be almost temperature independent and were treated as common parameters in the temperature range between 2 and 23.5~K. The resulting values were obtained as $A_{\rm AF1}=0.0065$ and $A_{\rm AF2}=0.0811$ [Fig.~\ref{ZF_para}(d)].

\begin{figure}[t]
   \begin{center}
     \includegraphics[width=75mm]{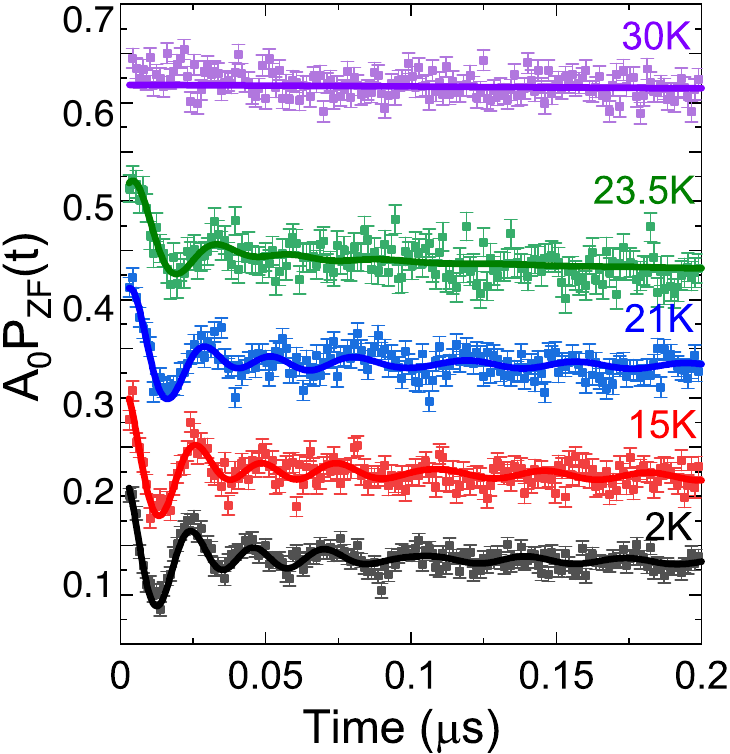}
   \end{center}
   \caption{Temperature-dependent \musr~spectra for \knto~taken under zero-field. Solid lines show the best fits with Eq.~\ref{2}. Each spectrum is offset along the y-axis by 0.1, for clarity of display.}
   \label{ZF_tem}
 \end{figure}

Figure~\ref{ZF_tem} shows the temperature dependence of ZF-\musr~time spectra [$t$ $<$ 0.2 $\mu$s] recorded at temperature between 2 and 30 K. The time spectra recorded below \TN~(= 26 K) were well fitted using the Eq.~\ref{2} in both long and short time domains. 
Fig.~\ref{ZF_para} shows the temperature dependencies of the \musr~parameters obtained by fitting the ZF-\musr~spectrum with Eq.~\ref{2}. 
As temperature decreases from 40~K, both $f_{\rm AF1}$ and $f_{\rm AF2}$ drastically increase ($\sim$ 75 - 93 $\%$ of its base temperature value) at around 25~K [Fig.~\ref{ZF_para}(a)]. Since $f_{\rm AF}$ corresponds to the order parameter of a magnetic transition, such a rapid AF transition is most likely indication of a first-order transition, which could be linked to a multiple structural phase transition. However, the co-existence of a structural and magnetic transition needs to be further investigated by low-temperature x-ray/neutron diffraction. Furthermore, these two frequencies seem to abruptly disappear almost at the same temperature \TN~(= 26~K). This suggests that the two frequencies are not caused by the coexistence of two different phases in the sample but by two magnetically inequivalent muon sites in the lattice.
Although both $\lambda_{\rm AF1}$ and $\lambda_{\rm AF2}$ are roughly temperature independent at temperature below 20~K, $\lambda_{\rm AF2}$ increases with temperature below the vicinity of \TN, indicating the increase in the field fluctuation towards \TN~(Fig. \ref{ZF_para}(b)). 

$\phi_{\rm AF}$ is almost constant below 18~K, i.e. $\phi_{\rm AF}$ $\sim$ -20$^{\circ}$, while the magnitude of $\phi_{\rm AF}$ increases with temperature above 18~K [Fig.~\ref{ZF_para}(c)]. This suggests that the spin structure is most likely commensurate (C) to the crystal lattice. This is because an incommensurate (IC) AF structure usually provides a much large phase delay for a cosine function, typically -45 – -60$^{\circ}$, due to the mismatch between the IC magnetic modulation and muon sites. Indeed, usually a commensurate magnetic ordering gives $\phi_{\rm AF}\approx0$. The observed small phase delay (-20$^{\circ}$) could instead be related to an artificial effect from the fit of very initial time domain for the fast oscillation. It could also be an effect from multiple muon stopping sites \cite{Sugiyama2011,LiCrO2}. As an conclusion, the small delay of the initial phase is likely to support commensurate AF order in \knto, however, we would need further low-temperature neutron experiment to confirm this.



Finally, all the \musr~parameters under ZF show a monotonic change in the temperature range between 2~K and \TN. The present results hence suggest the absence of an additional magnetic transition down to 2~K, which is in good agreement with the magnetisation and wTF-\musr~results. Additional neutron diffraction studies at low temperature would be the next future and natural step to shed further light on the magnetic nature of \knto.

\
\
\
\
\
\
\

 \begin{figure}[t]
   \begin{center}
     \includegraphics[width=75mm]{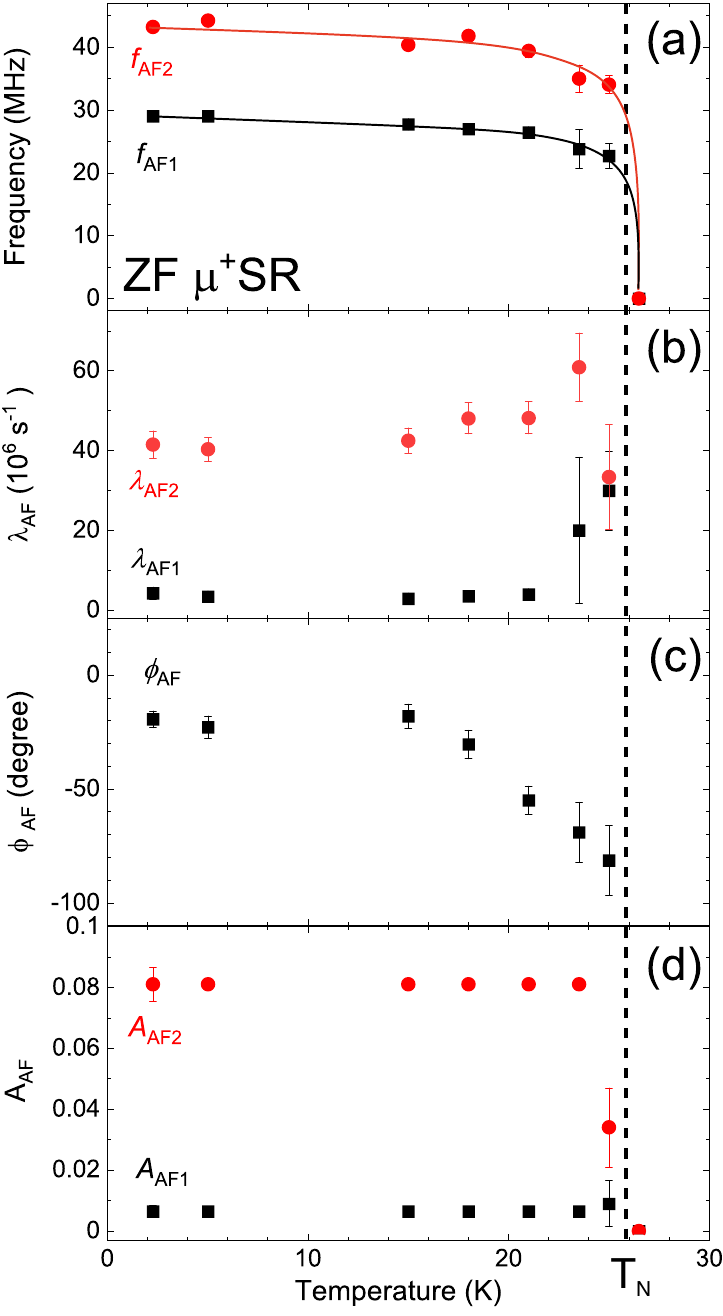}
   \end{center}
  \caption{Temperature dependencies of the ZF-\musr~parameters for \knto; (a) muon-spin precession frequencies (f$_{\rm AF1}$ and f$_{\rm AF2}$), lines are guide to the eyes. (b) the relaxation rates ($\lambda_{AF}$), (c) the common initial phases of the two oscillatory signals ($\phi_{\rm AF}$) and (d) the asymmetries (A$_{\rm AF}$). The data were obtained by fitting the ZF-\musr~spectra using Eq. \ref{2}. Vertical dashed line indicates the antiferromagnetic transition temperature $T_{\rm N}\approx26$~K determined by wTF-\musr~measurments.}
   \label{ZF_para}
 \end{figure} 

 \subsection{\label{HT_MUSR} High-temperature K-ion Diffusive Properties} 
To study the solid-state K-ion diffusive properties of \knto, \musr~measurements above the magnetic transition temperature were performed. Both Li-ion \cite{Sugiyama2011, Sugiyama2009, Sugiyama2012} and Na-ion \cite{Mansson2013, Umegaki2018} diffusive properties as a function of temperature have been already extensively studied using a series of ZF, wTF and LF-\musr~time spectra measurements, where LF means the applied field parallel to the initial muon spin polarization. 
However, since the nuclear magnetic moment of K ($\mu$[$^{39}$K] = 0.39 $\mu_{\rm N}$) is much smaller than that of Li ($\mu$[$^7$Li] = 3.26 $\mu_{\rm N}$) and Na ($\mu$[$^{23}$Na] = 2.22 $\mu_{\rm N}$), the measurement of K-ion dynamics using microscopic magnetic techniques \cite{Alloul2012, Siegel2001} is challenging. This means that \musr~could provide unique information on the K-ion diffusive properties, through its sensitivity to local magnetic environments.  


 \begin{figure}[ht]
   \begin{center}
     \includegraphics[width=75mm]{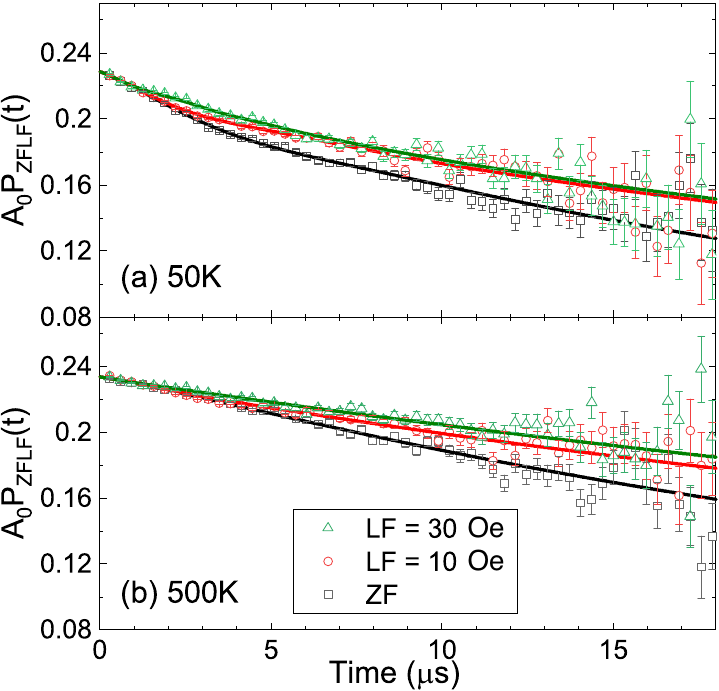}
   \end{center}
   \caption{ZF and two LF (10 and 30 Oe) \musr~time spectra measured at (a) 50 K and (b) 500 K. Solid lines represent the fit result using Eq. \ref{3}}
   \label{wTF_dif}
 \end{figure}
 

In what follows, \musr~time spectra were collected in the temperature range between 50 and 550~K using the EMU instrument of ISIS in UK. Fig. \ref{wTF_dif} shows the ZF- and LF-\musr~time spectrum obtained at 50~K and 500~K. A decoupling behaviour by the applied LF (= 10 and 30~Oe), i.e. a reduction in the relaxation rate, is seen even for small fields at both temperatures, suggesting that $H_{\rm int}$ is mainly formed by nuclear magnetic moments. The small nuclear moment of each element (K, Ni, Te and O) in the compound yelds small widths of the field distributions at the muon sites (small $\Delta$), resulting in the almost exponentially relaxing spectrum. This is also why it is essential to conduct these measurements at a pulsed muon facility that gives access to a longer time domain, and thereby yields a more robust fit to the data.

At each temperature, the ZF and the two LF spectra were fitted by a combination of two exponentially relaxing dynamic Gaussian Kubo-Toyabe (KT) functions. This would then be the same two muon sites also observed in the low-temperatrure \musr~ study (Eq. \ref{2} and Fig. \ref{ZF_para}(a)). In addition, there is a background signal from the fraction of muons stopped mainly in the silver plate mask on the sample holder. The resulting fit function for the ZF and two LF spectra is as follows:



\begin{eqnarray}
A_0 \, P_{\rm LF}(t) &=& A_{\rm KT1}G^{\rm DGKT}(\Delta _1, v_1, t, H_{\rm LF}) \times \exp({-\lambda_{\rm KT1}t})
\cr
&+&  A_{\rm KT2}G^{\rm DGKT}(\Delta _2, v_2, t, H_{\rm LF}) \times \exp({-\lambda_{\rm KT2}t})
\cr
&+& A_{\rm BG},
\label{3}
\end{eqnarray}

where $A_0$ is the initial asymmetry, $A_{\rm KT1}$, $A_{\rm KT2}$ and $A_{\rm BG}$ are the asymmetries associated with the three signals, $\Delta_1$ and $\Delta_2$ are related with the width of the local field distributions at the muon sites, and $\nu_1$ and $\nu_2$ are the field fluctuation rates and $\lambda_{\rm KT1}$ and $\lambda_{\rm KT2}$ are the relaxation rates. When $\nu$ = 0  and $H_{\rm LF1}$ = 0, $G^{\rm DGKT}(\Delta _1, v_1, t, H_{\rm LF1})$ is a static Gaussian KT function in ZF. Furthermore, a fitting procedure with a common background asymmetry ($A_{BG}$ $\sim$ 0.04723) was employed over the entire temperature range, but with a temperature dependent $\nu$ and $\lambda$ (see also Fig. \textcolor{blue}{S2}). $\Delta$ was also treated as a common parameter over the entire temperature range [$\Delta_1$ $\sim$ 0.291 (23) $\mu$s $^{-1}$ and $\Delta_2$ $\sim$ 0.043 (9) $\mu$s $^{-1}$]. Since $\Delta_{\rm KT}$ are one order of magnitude different, this could be the result of the existance of two distinct muon sites with different local field distributions, which is consistent with the two frequencies (two muon sites) observed at low-temperature. 



 \begin{figure}[h]
   \begin{center}
     \includegraphics[width=75mm]{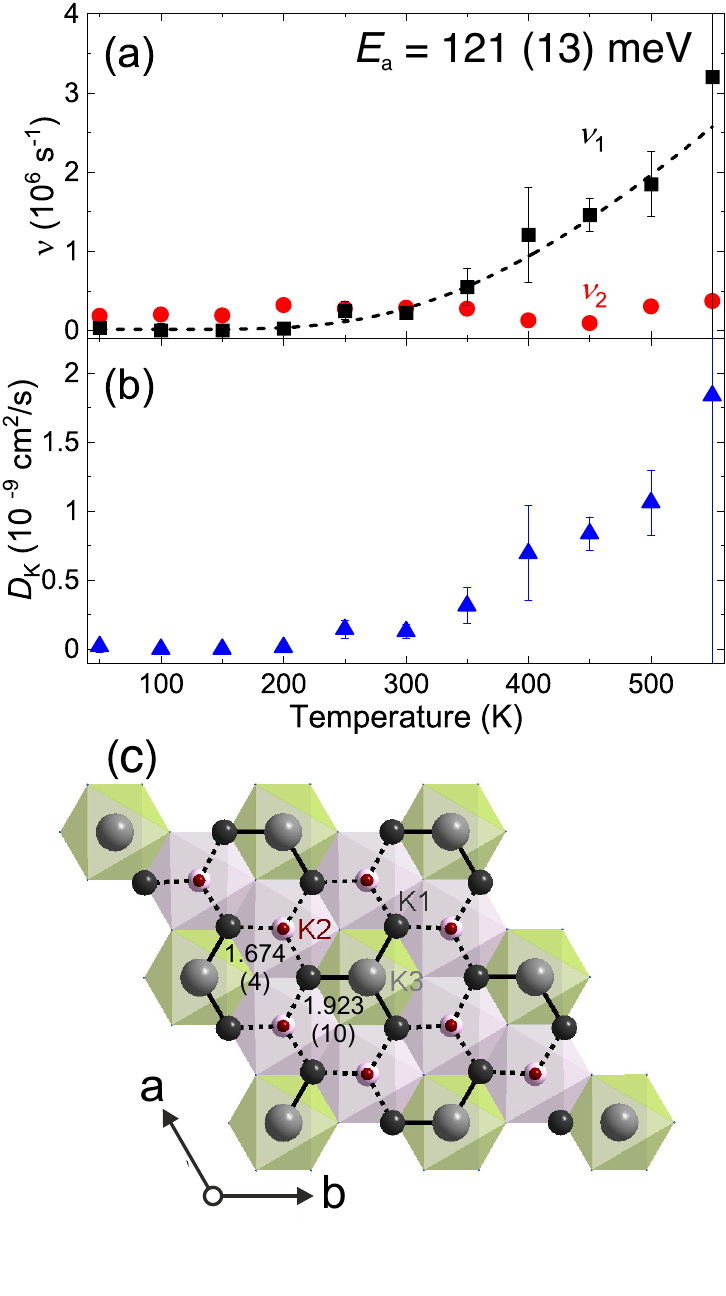}
  \end{center}
   \caption{The temperature dependencies of (a) $\nu$ and (b) diffusion coefficient for \knto. The black dashed line is fit to an Arrhenius equation $\nu$ = $A \times {\rm exp}(-E_{\rm a}/k_{\rm B}T)$, which yield an activation energy of $E_{\rm a}$ = 121 (13)~meV. Each data point was obtained by fitting the ZF and LF (= 10 and 30 Oe) spectra using Eq. \ref{3}. (c) Crystal structure of \knto~projection along \cc-axis. Diffusion paths, K1 - K2 ($s_1$ = 1.674 \AA) and K1 - K3 ($s_2$ = 1.923 \AA), are illustrated by dot line and solid line, respectively.} 
   \label{diffu_para}
\end{figure}



Fig. \ref{diffu_para}(a) shows the temperature dependencies of $\nu_1$ and $\nu_2$. $\nu_2$ is almost constant over the whole temperature range ($\nu_2$ at 50~K is 0.184 (31)), while $\nu_1$ is close to zero up to 200~K, and then starts to clearly increase from 250~K. The increase of $\nu_1$ between 250 and 550~K is explained by a thermal activation process, which signals the onset of diffusive motion of either \ce{K^+} or $\mu^+$ above 250~K. Here, the scenario of K-ion diffusion is strongly supported by electrochemical investigations that clearly indicate that the K-ions are mobile in this temperature range \cite{Titus2020, Masese2018}. Assuming the K-diffusion in \knto, the fit of the $\nu_1$($T$) curve with an Arrhenius type equation (dashed line in Fig. \ref{diffu_para}(a)) provides the activation energy as $E_{\rm a}$ = 121 (13)~meV. This value is comparable to the activation energy obtained for Li based layered oxides by \musr, e.g. $E_{\rm a}=96$~meV for \ce{Li_{0.53}CoO2} \cite{Sugiyama2009}, $E_{\rm a}=124$~meV for \ce{Li_{0.98}Ni_{1.02}O2} \cite{Mansson2013}.


We attempt to estimate a diffusion coefficient of the K-ions ($D_{\rm K}$) using the obtained fluctuation rate $\nu$ as a direct measure of the ion hopping rate. The field fluctuation rate has, in the past, been successfully used to determine the diffusion coefficient for Li and Na compounds \cite{Sugiyama2009, Umegaki2018}. The principle of diffusion of K$^+$ should be naturally the same to those for Li$^+$ and Na$^+$. Consequently, $D_{\rm K}$ is estimated via:

\begin{eqnarray}
D_{\rm K} &=& \sum_{i=1}^{h} \frac{1}{N_i} Z_{\rm v,i} s_i^2 \nu,
\label{4}
\end{eqnarray}

where $N$ is the number of K sites, $Z$ is the vacancy fraction and $s$ is the jump distance. Naturally, we restrict the diffusion path within the 2D layer of the honeycomb. Moreover, we assume a diffusion path only within the nearest neighbour sites within the honeycomb flower as shown in Fig.~\ref{diffu_para} (c) where only two K-diffusion pathways are allowed, that is, K1 to K2 and K1 to K3. The values for $s$ and $Z$ are extracted from our neutron diffraction measurements [see also the refined structural parameters in Tables. \textcolor{blue}{S1 and S2}], Since $s$ directly relates to the inter atomic distances of potassium, $s_1=1.674$ \AA~for K1 to K2 ($N_1=5$) and $s_2=1.923$ \AA~for K1 to K3 ($N_2= 4$). Based on such assumption, $D_{\rm K}$ is estimated as 0.13 $\times$ 10$^{-9}$ cm$^2$/s for $\nu_1$(300K) = 0.29 $\mu$s$^{-1}$. This value is lower by one order of magnitude  than $D_{\rm K}$ of \ce{LiCoO2} \cite{Sugiyama2009}. $D_{\rm K}$ are also calculated for the other temperatures as shown in Fig. \ref{diffu_para}(b), e.g. $D_{\rm K}$ = 0.69 $\times$ 10$^{-9}$ cm$^2$/s using $\nu_1$(400K) = 1.21 $\mu$s$^{-1}$ and 1.06 $\times$ 10$^{-9}$ cm$^2$/s using $\nu_1$(500K) = 1.85 $\mu$s$^{-1}$. Here we also assumed that the atomic structure remains the same within the entire temperature range. To further investigate the ion diffusion in \knto, detailed studies of the temperature dependency of the atomic structure using x-ray and/or neutron diffraction would be needed. Such investigations could yield even more accurate information on the active diffusion pathways \cite{Medarde2013}, which also allow us to further refine the calculations of $D_{\rm K}$ from $\nu(T)$, especially as a function of temperature.

\section{\label{sec:discussion}Discussion}
Finally, it should be noted that the KT1 signal that reveals the strong temperature dependence in K-ion hopping rate ($\nu_{1}$) constitutes the minor volume fraction (asymmetry). This could be due to that the two different muon stopping sites are very different in relation to the K-ion layers, and that KT2 is related to a site where the muon is screened from detecting dynamic changes in the weak nuclear moment of potassium. Such scenario is supported by the fact that in the low-temperature \musr~data the larger volume fraction relates to the higher frequency ($f_{\rm AF2}$), which indicate that such muon site is located closer to the TMO layer. Another, probable scenario is that KT1 and KT2 relates to surface and bulk signals, respectively. In our previous work on the well-known \ce{LiFePO4} cathode material \cite{Sugiyama2011, Sugiyama2012, Benedek2019,Benedek2020} we have shown that the self-diffusion of lithium ions is mainly limited to the surface region of the cathode material particles. The current results from \knto~indicate that it could be possible to improve K-ion diffusion and thereby device performance by reducing the particle size (i.e. larger surface area). Consequently, further \musr~studies of nano-particles of \knto~will be of very high interest.

As a conclusion, muon spin rotation and relaxation (\musr) together with magnetization measurements on \knto~reveal the onset of commensurate-like antiferromagnetic transitions at 26 K. Further, potassium-ions (K$^+$) in \knto~have been found to be mobile above 250~K, with remarkably low activation energy of 121 (13)~meV which is comparable to thermal activation energy scales of related lithium- and sodium-based materials. Moreover, we also estimate a diffusion coefficient of K-ion ($D_{\rm K}$) as a function of temperature. This brings related honeycomb layered oxide materials to the foreground of fast ionic conductors for energy storage. Moreover, we have shown, for the first time, the feasibility of \musr~study in investigating the K$^+$ dynamics of materials which tend to have low nuclear magnetic moments. This study expands the research frontier of alkali-ion dynamics within materials, previously relegated mainly to lithium and sodium. Finally our results also indicate an interesting possibility that K-ion self diffusion at the surface of the powder particles are dominating. This opens future possibilities for improving ion diffusion and device performance using nano-structuring.

\section{\label{sec:exp}Experimental Section}

\subsection{\label{sec:level2} Materials synthesis} 
Polycrystalline powder of \knto, (\ce{K_{2/3}Ni_{2/3}Te_{1/3}O2}) was synthesised using a high-temperature ceramics route. Stoichiometric amounts of \ce{NiO} (99.9 $\%$ purity, Kojundo Chemical Laboratory (Japan)), \ce{TeO2} (99.0 $\%$ purity, Aldrich) and \ce{K2CO3} (99.9$\%$ purity, Rare Metallic (Japan)) were mixed, pressed into pellets and finally heated for 23 h at 800\doc~in air. The obtained powders were stored in an argon-purged glove box that was maintained at a dew point of below - 80\doc~dP, to prevent exposure of the materials to moisture. More detailed information on the synthesis protocol can be found in Ref.\cite{Masese2018}.

\subsection{\label{sec:level2} X-ray and Neutron Powder Diffraction}
Sample quality was checked by room-temperature x-ray powder diffraction (Cu-K$\alpha$ radiation). 
Room-temperature neutron powder diffraction was performed on the high-resolution time-of-flight SPICA beamline at J-PARC \cite{Yonemura2014}. Structural refinements were performed with the FullProf suite of programs \cite{Rodriguez-Carvajal1993a}, taking into account anisotropic strains using a spherical harmonics modelling of the Bragg peak broadening. For the diffraction experiments the samples were carefully packed and sealed inside a glove-box to avoid sample degradation or contamination. 

\subsection{\label{sec:level2} Magnetic susceptibility measurements}
Magnetic measurements as a function of temperature were performed with a 9~T Quantum Design superconducting quantum interference device (SQUID) magnetometer in zero-field-cooled ($zfc$) and field-cooled ($fc$) modes upon heating $T=5-300$~K. The magnetic susceptibility ($\chi$) was obtained based on the equation $\chi$ = $M$/$H$, where $M$ is the magnetisation obtained by dividing the measured magnetic moment by the sample mass and $H$ is the external applied magnetic field (in Oe).

\subsection{\label{sec:level2}Muon spin rotation and relaxation (\musr) measurements}

\musr~experiments were performed at a muon beam line using the surface muon spectrometer GPS at the Swiss Muon Source (S$\mu$S) of PSI in Switzerland. The powder sample ($m\approx0.5$~g) was placed and sealed inside an envelope made out of a 25~$\mu$m-thick silver foil covering a surface area of 1 $\times$ 1 cm$^2$. The sample preparation was performed in a helium glove-box to avoid sample degradation. In order to minimise the background signal, an envelope was attached to a fork-type Cu sample holder using a single layer of Al-coated mylar tape. The sample holder was attached to a stick and inserted into a liquid-He flow-type cryostat for measurements in the temperature range of 2 - 50~K where weak transverse-field (wTF) of 20~Oe, and zero-field (ZF) \musr~time spectra were collected.

 \begin{figure*}[ht]
   \begin{center}
     \includegraphics[width=170mm]{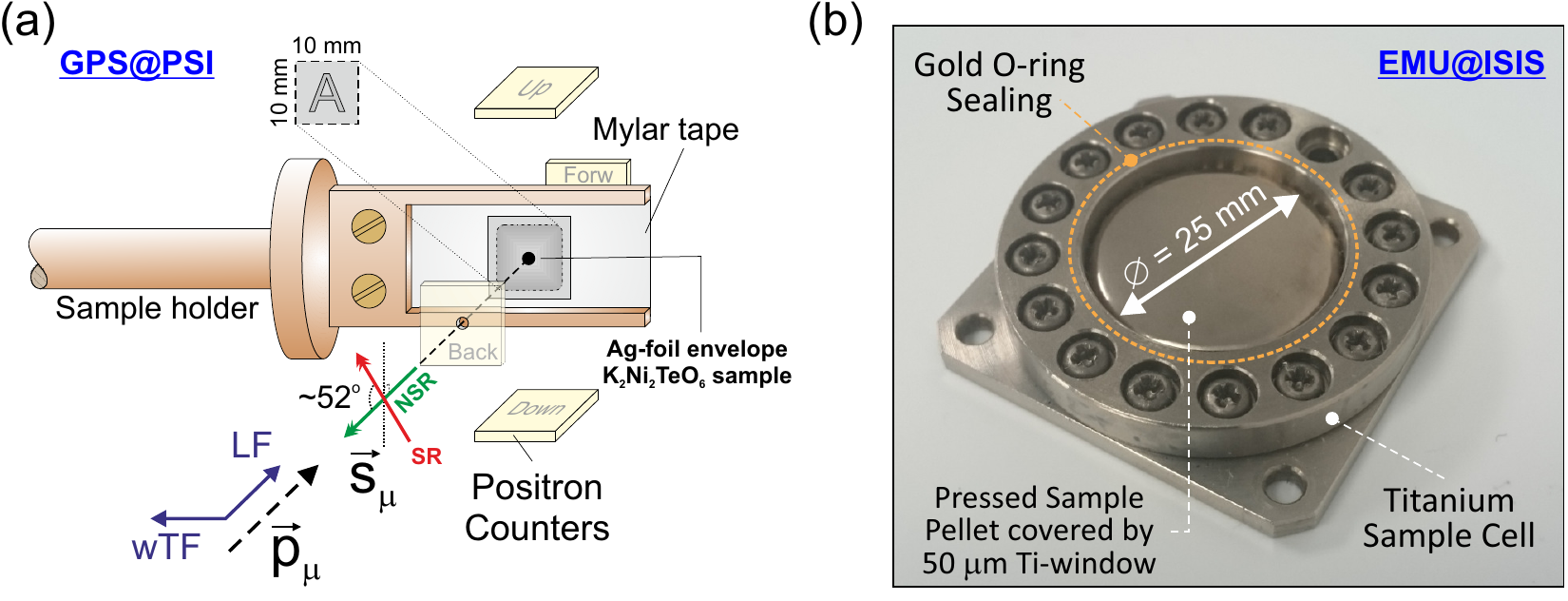}
   \end{center}
   \caption{(a) Illustration of the experimental setup used to perform muon spin rotation and relaxation (\musr) measurements at the GPS/PSI spectrometer. The muons are implanted into the sample through the back detector, where $\vec{P}_{\mu}$ is the momentum vector of the muon and $\vec{S}_{\mu}$ the spin vector pointing away from the direction of motion. The muons stops inside the sample and subsequently decay into positrons, which are counted by the detectors, and neutrinos. LF denotes the longitudinal field whereas wTF is the weak transverse field applied. (b) Sealed and mounted titanium sample cell used for the \musr~ion diffusion measurements at EMU/ISIS.}
   \label{GPSEMU}
 \end{figure*} 

For the high-temperature ion diffusion measurements, \musr~time spectra were recorded using the EMU spectrometer at the pulsed muon source of ISIS/RAL in UK. A powder sample of \knto~($m\approx$ 1~g) was pressed into a pellet with a diameter and thickness of 25~mm and 3.0~mm, respectively. This pellet was packed into a sealed (gold O-ring) powder cell made of non-magnetic titanium using a thin (50~$\mu$m) Ti-film window. Sample preparation was performed inside a helium glove-box to avoid sample degradation. In addition, a silver mask was mounted onto the Ti-cell to ensure that there would be no non-relaxing minor background signal in a wide temperature range. The cell was mounted onto a Cu end-plate of the closed-cycle refrigerator (CCR) and measurements were performed at temperatures between 50 and 550 K. \musr~time spectra were subsequently collected at ZF, wTF = 20~Oe and longitudinal field (LF = 10 and 30~Oe).

Further details regarding the experimental techniques and set-ups are provided in Fig.~\ref{GPSEMU} as well as in Ref.~\cite{Yaouanc2011}. The musrfit \cite{Suter2012} software package was used to analyse the \musr~data.

\bibliography{library} 

\begin{thebibliography}{36}%
\makeatletter
\providecommand \@ifxundefined [1]{%
 \@ifx{#1\undefined}
}%
\providecommand \@ifnum [1]{%
 \ifnum #1\expandafter \@firstoftwo
 \else \expandafter \@secondoftwo
 \fi
}%
\providecommand \@ifx [1]{%
 \ifx #1\expandafter \@firstoftwo
 \else \expandafter \@secondoftwo
 \fi
}%
\providecommand \natexlab [1]{#1}%
\providecommand \enquote  [1]{``#1''}%
\providecommand \bibnamefont  [1]{#1}%
\providecommand \bibfnamefont [1]{#1}%
\providecommand \citenamefont [1]{#1}%
\providecommand \href@noop [0]{\@secondoftwo}%
\providecommand \href [0]{\begingroup \@sanitize@url \@href}%
\providecommand \@href[1]{\@@startlink{#1}\@@href}%
\providecommand \@@href[1]{\endgroup#1\@@endlink}%
\providecommand \@sanitize@url [0]{\catcode `\\12\catcode `\$12\catcode
  `\&12\catcode `\#12\catcode `\^12\catcode `\_12\catcode `\%12\relax}%
\providecommand \@@startlink[1]{}%
\providecommand \@@endlink[0]{}%
\providecommand \url  [0]{\begingroup\@sanitize@url \@url }%
\providecommand \@url [1]{\endgroup\@href {#1}{\urlprefix }}%
\providecommand \urlprefix  [0]{URL }%
\providecommand \Eprint [0]{\href }%
\providecommand \doibase [0]{http://dx.doi.org/}%
\providecommand \selectlanguage [0]{\@gobble}%
\providecommand \bibinfo  [0]{\@secondoftwo}%
\providecommand \bibfield  [0]{\@secondoftwo}%
\providecommand \translation [1]{[#1]}%
\providecommand \BibitemOpen [0]{}%
\providecommand \bibitemStop [0]{}%
\providecommand \bibitemNoStop [0]{.\EOS\space}%
\providecommand \EOS [0]{\spacefactor3000\relax}%
\providecommand \BibitemShut  [1]{\csname bibitem#1\endcsname}%
\let\auto@bib@innerbib\@empty
\bibitem [{\citenamefont {Masese}\ \emph {et~al.}(2018)\citenamefont {Masese},
  \citenamefont {Yoshii}, \citenamefont {Yamaguchi}, \citenamefont {Okumura},
  \citenamefont {Huang}, \citenamefont {Kato}, \citenamefont {Kubota},
  \citenamefont {Furutani}, \citenamefont {Orikasa}, \citenamefont {Senoh},
  \citenamefont {Sakaebe},\ and\ \citenamefont {Shikano}}]{Masese2018}%
  \BibitemOpen
  \bibfield  {author} {\bibinfo {author} {\bibfnamefont {T.}~\bibnamefont
  {Masese}}, \bibinfo {author} {\bibfnamefont {K.}~\bibnamefont {Yoshii}},
  \bibinfo {author} {\bibfnamefont {Y.}~\bibnamefont {Yamaguchi}}, \bibinfo
  {author} {\bibfnamefont {T.}~\bibnamefont {Okumura}}, \bibinfo {author}
  {\bibfnamefont {Z.~D.}\ \bibnamefont {Huang}}, \bibinfo {author}
  {\bibfnamefont {M.}~\bibnamefont {Kato}}, \bibinfo {author} {\bibfnamefont
  {K.}~\bibnamefont {Kubota}}, \bibinfo {author} {\bibfnamefont
  {J.}~\bibnamefont {Furutani}}, \bibinfo {author} {\bibfnamefont
  {Y.}~\bibnamefont {Orikasa}}, \bibinfo {author} {\bibfnamefont
  {H.}~\bibnamefont {Senoh}}, \bibinfo {author} {\bibfnamefont
  {H.}~\bibnamefont {Sakaebe}}, \ and\ \bibinfo {author} {\bibfnamefont
  {M.}~\bibnamefont {Shikano}},\ }\href {\doibase 10.1038/s41467-018-06343-6}
  {\bibfield  {journal} {\bibinfo  {journal} {Nature Communications}\ }\textbf
  {\bibinfo {volume} {9}},\ \bibinfo {pages} {3823} (\bibinfo {year}
  {2018})}\BibitemShut {NoStop}%
\bibitem [{\citenamefont {Matsubara}\ \emph {et~al.}(2019)\citenamefont
  {Matsubara}, \citenamefont {Petit}, \citenamefont {Martin}, \citenamefont
  {Fauth}, \citenamefont {Suard}, \citenamefont {Rols},\ and\ \citenamefont
  {Damay}}]{Matsubara2019c}%
  \BibitemOpen
  \bibfield  {author} {\bibinfo {author} {\bibfnamefont {N.}~\bibnamefont
  {Matsubara}}, \bibinfo {author} {\bibfnamefont {S.}~\bibnamefont {Petit}},
  \bibinfo {author} {\bibfnamefont {C.}~\bibnamefont {Martin}}, \bibinfo
  {author} {\bibfnamefont {F.}~\bibnamefont {Fauth}}, \bibinfo {author}
  {\bibfnamefont {E.}~\bibnamefont {Suard}}, \bibinfo {author} {\bibfnamefont
  {S.}~\bibnamefont {Rols}}, \ and\ \bibinfo {author} {\bibfnamefont
  {F.}~\bibnamefont {Damay}},\ }\href {\doibase 10.1103/PhysRevB.100.220406}
  {\bibfield  {journal} {\bibinfo  {journal} {Physical Review B}\ }\textbf
  {\bibinfo {volume} {100}},\ \bibinfo {pages} {220406(R)} (\bibinfo {year}
  {2019})}\BibitemShut {NoStop}%
\bibitem [{\citenamefont {Lee}\ \emph {et~al.}(2017)\citenamefont {Lee},
  \citenamefont {Wang}, \citenamefont {Zhao}, \citenamefont {Li}, \citenamefont
  {Lynn}, \citenamefont {Harris}, \citenamefont {Rule}, \citenamefont {Yang},\
  and\ \citenamefont {Berger}}]{Lee2017}%
  \BibitemOpen
  \bibfield  {author} {\bibinfo {author} {\bibfnamefont {C.-H.}\ \bibnamefont
  {Lee}}, \bibinfo {author} {\bibfnamefont {C.-W.}\ \bibnamefont {Wang}},
  \bibinfo {author} {\bibfnamefont {Y.}~\bibnamefont {Zhao}}, \bibinfo {author}
  {\bibfnamefont {W.-H.}\ \bibnamefont {Li}}, \bibinfo {author} {\bibfnamefont
  {J.~W.}\ \bibnamefont {Lynn}}, \bibinfo {author} {\bibfnamefont {A.~B.}\
  \bibnamefont {Harris}}, \bibinfo {author} {\bibfnamefont {K.}~\bibnamefont
  {Rule}}, \bibinfo {author} {\bibfnamefont {H.-D.}\ \bibnamefont {Yang}}, \
  and\ \bibinfo {author} {\bibfnamefont {H.}~\bibnamefont {Berger}},\ }\href
  {\doibase 10.1038/s41598-017-06651-9} {\bibfield  {journal} {\bibinfo
  {journal} {Scientific Reports}\ }\textbf {\bibinfo {volume} {7}},\ \bibinfo
  {pages} {6437} (\bibinfo {year} {2017})}\BibitemShut {NoStop}%
\bibitem [{\citenamefont {Karna}\ \emph {et~al.}(2017)\citenamefont {Karna},
  \citenamefont {Zhao}, \citenamefont {Sankar}, \citenamefont {Avdeev},
  \citenamefont {Tseng}, \citenamefont {Wang}, \citenamefont {Shu},
  \citenamefont {Matan}, \citenamefont {Guo},\ and\ \citenamefont
  {Chou}}]{Karna2017}%
  \BibitemOpen
  \bibfield  {author} {\bibinfo {author} {\bibfnamefont {S.~K.}\ \bibnamefont
  {Karna}}, \bibinfo {author} {\bibfnamefont {Y.}~\bibnamefont {Zhao}},
  \bibinfo {author} {\bibfnamefont {R.}~\bibnamefont {Sankar}}, \bibinfo
  {author} {\bibfnamefont {M.}~\bibnamefont {Avdeev}}, \bibinfo {author}
  {\bibfnamefont {P.~C.}\ \bibnamefont {Tseng}}, \bibinfo {author}
  {\bibfnamefont {C.~W.}\ \bibnamefont {Wang}}, \bibinfo {author}
  {\bibfnamefont {G.~J.}\ \bibnamefont {Shu}}, \bibinfo {author} {\bibfnamefont
  {K.}~\bibnamefont {Matan}}, \bibinfo {author} {\bibfnamefont {G.~Y.}\
  \bibnamefont {Guo}}, \ and\ \bibinfo {author} {\bibfnamefont {F.~C.}\
  \bibnamefont {Chou}},\ }\href {\doibase 10.1103/PhysRevB.95.104408}
  {\bibfield  {journal} {\bibinfo  {journal} {Physical Review B}\ }\textbf
  {\bibinfo {volume} {95}},\ \bibinfo {pages} {104408} (\bibinfo {year}
  {2017})}\BibitemShut {NoStop}%
\bibitem [{\citenamefont {Kim}\ \emph {et~al.}(2016)\citenamefont {Kim},
  \citenamefont {Deng}, \citenamefont {Li}, \citenamefont {Gupta},
  \citenamefont {Akamatsu}, \citenamefont {Gopalan},\ and\ \citenamefont
  {Greenblatt}}]{Kim2016a}%
  \BibitemOpen
  \bibfield  {author} {\bibinfo {author} {\bibfnamefont {S.~W.}\ \bibnamefont
  {Kim}}, \bibinfo {author} {\bibfnamefont {Z.}~\bibnamefont {Deng}}, \bibinfo
  {author} {\bibfnamefont {M.-R.}\ \bibnamefont {Li}}, \bibinfo {author}
  {\bibfnamefont {A.~S.}\ \bibnamefont {Gupta}}, \bibinfo {author}
  {\bibfnamefont {H.}~\bibnamefont {Akamatsu}}, \bibinfo {author}
  {\bibfnamefont {V.}~\bibnamefont {Gopalan}}, \ and\ \bibinfo {author}
  {\bibfnamefont {M.}~\bibnamefont {Greenblatt}},\ }\href
  {http://pubs.acs.org/doi/abs/10.1021/acs.inorgchem.5b02677{\#}.WJ3ec-cNw8o.mendeley}
  {\bibfield  {journal} {\bibinfo  {journal} {Inorganic Chemistry}\ }\textbf
  {\bibinfo {volume} {55}},\ \bibinfo {pages} {1333} (\bibinfo {year}
  {2016})}\BibitemShut {NoStop}%
\bibitem [{\citenamefont {Yang}\ \emph {et~al.}(2017)\citenamefont {Yang},
  \citenamefont {Jiang}, \citenamefont {Deng}, \citenamefont {Wang},
  \citenamefont {Chen},\ and\ \citenamefont {Huang}}]{Yang2017a}%
  \BibitemOpen
  \bibfield  {author} {\bibinfo {author} {\bibfnamefont {Z.}~\bibnamefont
  {Yang}}, \bibinfo {author} {\bibfnamefont {Y.}~\bibnamefont {Jiang}},
  \bibinfo {author} {\bibfnamefont {L.}~\bibnamefont {Deng}}, \bibinfo {author}
  {\bibfnamefont {T.}~\bibnamefont {Wang}}, \bibinfo {author} {\bibfnamefont
  {S.}~\bibnamefont {Chen}}, \ and\ \bibinfo {author} {\bibfnamefont
  {Y.}~\bibnamefont {Huang}},\ }\href {\doibase 10.1016/j.jpowsour.2017.06.014}
  {\bibfield  {journal} {\bibinfo  {journal} {Journal of Power Sources}\
  }\textbf {\bibinfo {volume} {360}},\ \bibinfo {pages} {319} (\bibinfo {year}
  {2017})}\BibitemShut {NoStop}%
\bibitem [{\citenamefont {Khanh}\ \emph {et~al.}(2016)\citenamefont {Khanh},
  \citenamefont {Abe}, \citenamefont {Sagayama}, \citenamefont {Nakao},
  \citenamefont {Hanashima}, \citenamefont {Kiyanagi}, \citenamefont
  {Tokunaga},\ and\ \citenamefont {Arima}}]{Khanh2016}%
  \BibitemOpen
  \bibfield  {author} {\bibinfo {author} {\bibfnamefont {N.~D.}\ \bibnamefont
  {Khanh}}, \bibinfo {author} {\bibfnamefont {N.}~\bibnamefont {Abe}}, \bibinfo
  {author} {\bibfnamefont {H.}~\bibnamefont {Sagayama}}, \bibinfo {author}
  {\bibfnamefont {A.}~\bibnamefont {Nakao}}, \bibinfo {author} {\bibfnamefont
  {T.}~\bibnamefont {Hanashima}}, \bibinfo {author} {\bibfnamefont
  {R.}~\bibnamefont {Kiyanagi}}, \bibinfo {author} {\bibfnamefont
  {Y.}~\bibnamefont {Tokunaga}}, \ and\ \bibinfo {author} {\bibfnamefont
  {T.}~\bibnamefont {Arima}},\ }\href {\doibase 10.1103/PhysRevB.93.075117}
  {\bibfield  {journal} {\bibinfo  {journal} {Physical Review B}\ }\textbf
  {\bibinfo {volume} {93}},\ \bibinfo {pages} {075117} (\bibinfo {year}
  {2016})}\BibitemShut {NoStop}%
\bibitem [{\citenamefont {Chaudhary}\ \emph {et~al.}(2018)\citenamefont
  {Chaudhary}, \citenamefont {Srivastava},\ and\ \citenamefont
  {Patnaik}}]{Chaudhary2018}%
  \BibitemOpen
  \bibfield  {author} {\bibinfo {author} {\bibfnamefont {S.}~\bibnamefont
  {Chaudhary}}, \bibinfo {author} {\bibfnamefont {P.}~\bibnamefont
  {Srivastava}}, \ and\ \bibinfo {author} {\bibfnamefont {S.}~\bibnamefont
  {Patnaik}},\ }\href {\doibase 10.1063/1.5029115} {\bibfield  {journal}
  {\bibinfo  {journal} {AIP Conference Proceedings}\ }\textbf {\bibinfo
  {volume} {1942}},\ \bibinfo {pages} {130045} (\bibinfo {year}
  {2018})}\BibitemShut {NoStop}%
\bibitem [{\citenamefont {Choi}\ \emph {et~al.}(2019)\citenamefont {Choi},
  \citenamefont {Manni}, \citenamefont {Singleton}, \citenamefont {Topping},
  \citenamefont {Lancaster}, \citenamefont {Blundell}, \citenamefont {Adroja},
  \citenamefont {Zapf}, \citenamefont {Gegenwart},\ and\ \citenamefont
  {Coldea}}]{Choi2019}%
  \BibitemOpen
  \bibfield  {author} {\bibinfo {author} {\bibfnamefont {S.}~\bibnamefont
  {Choi}}, \bibinfo {author} {\bibfnamefont {S.}~\bibnamefont {Manni}},
  \bibinfo {author} {\bibfnamefont {J.}~\bibnamefont {Singleton}}, \bibinfo
  {author} {\bibfnamefont {C.~V.}\ \bibnamefont {Topping}}, \bibinfo {author}
  {\bibfnamefont {T.}~\bibnamefont {Lancaster}}, \bibinfo {author}
  {\bibfnamefont {S.~J.}\ \bibnamefont {Blundell}}, \bibinfo {author}
  {\bibfnamefont {D.~T.}\ \bibnamefont {Adroja}}, \bibinfo {author}
  {\bibfnamefont {V.}~\bibnamefont {Zapf}}, \bibinfo {author} {\bibfnamefont
  {P.}~\bibnamefont {Gegenwart}}, \ and\ \bibinfo {author} {\bibfnamefont
  {R.}~\bibnamefont {Coldea}},\ }\href {\doibase 10.1103/PhysRevB.99.054426}
  {\bibfield  {journal} {\bibinfo  {journal} {Physical Review B}\ }\textbf
  {\bibinfo {volume} {99}},\ \bibinfo {pages} {054426} (\bibinfo {year}
  {2019})},\ \Eprint {http://arxiv.org/abs/1810.03212} {arXiv:1810.03212}
  \BibitemShut {NoStop}%
\bibitem [{\citenamefont {Wooldridge}\ \emph {et~al.}(2005)\citenamefont
  {Wooldridge}, \citenamefont {{Mck Paul}}, \citenamefont {Balakrishnan},\ and\
  \citenamefont {Lees}}]{Wooldridge2005}%
  \BibitemOpen
  \bibfield  {author} {\bibinfo {author} {\bibfnamefont {J.}~\bibnamefont
  {Wooldridge}}, \bibinfo {author} {\bibfnamefont {D.}~\bibnamefont {{Mck
  Paul}}}, \bibinfo {author} {\bibfnamefont {G.}~\bibnamefont {Balakrishnan}},
  \ and\ \bibinfo {author} {\bibfnamefont {M.~R.}\ \bibnamefont {Lees}},\
  }\href {\doibase 10.1088/0953-8984/17/4/013} {\bibfield  {journal} {\bibinfo
  {journal} {Journal of Physics Condensed Matter}\ }\textbf {\bibinfo {volume}
  {17}},\ \bibinfo {pages} {707} (\bibinfo {year} {2005})},\ \Eprint
  {http://arxiv.org/abs/0406513v2} {arXiv:0406513v2 [arXiv:cond-mat]}
  \BibitemShut {NoStop}%
\bibitem [{\citenamefont {Schaak}\ \emph {et~al.}(2003)\citenamefont {Schaak},
  \citenamefont {Klimczuk}, \citenamefont {Foo},\ and\ \citenamefont
  {Cava}}]{Schaak2003}%
  \BibitemOpen
  \bibfield  {author} {\bibinfo {author} {\bibfnamefont {R.~E.}\ \bibnamefont
  {Schaak}}, \bibinfo {author} {\bibfnamefont {T.}~\bibnamefont {Klimczuk}},
  \bibinfo {author} {\bibfnamefont {M.~L.}\ \bibnamefont {Foo}}, \ and\
  \bibinfo {author} {\bibfnamefont {R.~J.}\ \bibnamefont {Cava}},\ }\href
  {http://www.nature.com/doifinder/10.1038/nature01774{\%}5Cnpapers3://publication/doi/10.1038/nature01774}
  {\bibfield  {journal} {\bibinfo  {journal} {Nature}\ }\textbf {\bibinfo
  {volume} {424}},\ \bibinfo {pages} {527} (\bibinfo {year}
  {2003})}\BibitemShut {NoStop}%
\bibitem [{\citenamefont {Takada}\ \emph {et~al.}(2009)\citenamefont {Takada},
  \citenamefont {Onoda}, \citenamefont {Choi}, \citenamefont {Argyriou},
  \citenamefont {Izumi}, \citenamefont {Sakurai}, \citenamefont
  {Takayama-Muromachiand},\ and\ \citenamefont {Sasaki}}]{Takada2009}%
  \BibitemOpen
  \bibfield  {author} {\bibinfo {author} {\bibfnamefont {K.}~\bibnamefont
  {Takada}}, \bibinfo {author} {\bibfnamefont {M.}~\bibnamefont {Onoda}},
  \bibinfo {author} {\bibfnamefont {Y.-N.}\ \bibnamefont {Choi}}, \bibinfo
  {author} {\bibfnamefont {D.~N.}\ \bibnamefont {Argyriou}}, \bibinfo {author}
  {\bibfnamefont {F.}~\bibnamefont {Izumi}}, \bibinfo {author} {\bibfnamefont
  {H.}~\bibnamefont {Sakurai}}, \bibinfo {author} {\bibfnamefont
  {E.}~\bibnamefont {Takayama-Muromachiand}}, \ and\ \bibinfo {author}
  {\bibfnamefont {T.}~\bibnamefont {Sasaki}},\ }\href {\doibase
  10.1021/cm8031237} {\bibfield  {journal} {\bibinfo  {journal} {Chemistry of
  Materials}\ }\textbf {\bibinfo {volume} {21}},\ \bibinfo {pages} {3693}
  (\bibinfo {year} {2009})}\BibitemShut {NoStop}%
\bibitem [{\citenamefont {Sugiyama}\ \emph {et~al.}(2004)\citenamefont
  {Sugiyama}, \citenamefont {Brewer}, \citenamefont {Ansaldo}, \citenamefont
  {Itahara}, \citenamefont {Tani}, \citenamefont {Mikami}, \citenamefont
  {Mori}, \citenamefont {Sasaki}, \citenamefont {H{\'{e}}bert},\ and\
  \citenamefont {Maignan}}]{Sugiyama2004}%
  \BibitemOpen
  \bibfield  {author} {\bibinfo {author} {\bibfnamefont {J.}~\bibnamefont
  {Sugiyama}}, \bibinfo {author} {\bibfnamefont {J.~H.}\ \bibnamefont
  {Brewer}}, \bibinfo {author} {\bibfnamefont {E.~J.}\ \bibnamefont {Ansaldo}},
  \bibinfo {author} {\bibfnamefont {H.}~\bibnamefont {Itahara}}, \bibinfo
  {author} {\bibfnamefont {T.}~\bibnamefont {Tani}}, \bibinfo {author}
  {\bibfnamefont {M.}~\bibnamefont {Mikami}}, \bibinfo {author} {\bibfnamefont
  {Y.}~\bibnamefont {Mori}}, \bibinfo {author} {\bibfnamefont {T.}~\bibnamefont
  {Sasaki}}, \bibinfo {author} {\bibfnamefont {S.}~\bibnamefont
  {H{\'{e}}bert}}, \ and\ \bibinfo {author} {\bibfnamefont {A.}~\bibnamefont
  {Maignan}},\ }\href {\doibase 10.1103/PhysRevLett.92.017602} {\bibfield
  {journal} {\bibinfo  {journal} {Physical Review Letters}\ }\textbf {\bibinfo
  {volume} {92}},\ \bibinfo {pages} {017602} (\bibinfo {year} {2004})},\
  \Eprint {http://arxiv.org/abs/0310516} {arXiv:0310516 [cond-mat]}
  \BibitemShut {NoStop}%
\bibitem [{\citenamefont {Hertz}\ \emph {et~al.}(2008)\citenamefont {Hertz},
  \citenamefont {Huang}, \citenamefont {McQueen}, \citenamefont {Klimczuk},
  \citenamefont {Bos}, \citenamefont {Viciu},\ and\ \citenamefont
  {Cava}}]{Hertz2008}%
  \BibitemOpen
  \bibfield  {author} {\bibinfo {author} {\bibfnamefont {J.~T.}\ \bibnamefont
  {Hertz}}, \bibinfo {author} {\bibfnamefont {Q.}~\bibnamefont {Huang}},
  \bibinfo {author} {\bibfnamefont {T.}~\bibnamefont {McQueen}}, \bibinfo
  {author} {\bibfnamefont {T.}~\bibnamefont {Klimczuk}}, \bibinfo {author}
  {\bibfnamefont {J.~W.}\ \bibnamefont {Bos}}, \bibinfo {author} {\bibfnamefont
  {L.}~\bibnamefont {Viciu}}, \ and\ \bibinfo {author} {\bibfnamefont {R.~J.}\
  \bibnamefont {Cava}},\ }\href {\doibase 10.1103/PhysRevB.77.075119}
  {\bibfield  {journal} {\bibinfo  {journal} {Physical Review B}\ }\textbf
  {\bibinfo {volume} {77}},\ \bibinfo {pages} {075119} (\bibinfo {year}
  {2008})}\BibitemShut {NoStop}%
\bibitem [{\citenamefont {Alexander}\ \emph {et~al.}(2020)\citenamefont
  {Alexander}, \citenamefont {Allen}, \citenamefont {Atala}, \citenamefont
  {Bowen}, \citenamefont {Coley}, \citenamefont {Goodenough}, \citenamefont
  {Katsnelson}, \citenamefont {Koonin}, \citenamefont {Krenn}, \citenamefont
  {Madsen}, \citenamefont {M{\aa}nsson}, \citenamefont {Mauranyapin},
  \citenamefont {Melvin}, \citenamefont {Rasel}, \citenamefont {Reichl},
  \citenamefont {Yampolskiy}, \citenamefont {Yasskin}, \citenamefont
  {Zeilinger},\ and\ \citenamefont {Lidstr{\"{o}}m}}]{Mansson2020}%
  \BibitemOpen
  \bibfield  {author} {\bibinfo {author} {\bibfnamefont {G.~M.}\ \bibnamefont
  {Alexander}}, \bibinfo {author} {\bibfnamefont {R.~E.}\ \bibnamefont
  {Allen}}, \bibinfo {author} {\bibfnamefont {A.}~\bibnamefont {Atala}},
  \bibinfo {author} {\bibfnamefont {W.}~\bibnamefont {Bowen}}, \bibinfo
  {author} {\bibfnamefont {A.~A.}\ \bibnamefont {Coley}}, \bibinfo {author}
  {\bibfnamefont {J.~B.}\ \bibnamefont {Goodenough}}, \bibinfo {author}
  {\bibfnamefont {M.~I.}\ \bibnamefont {Katsnelson}}, \bibinfo {author}
  {\bibfnamefont {E.~V.}\ \bibnamefont {Koonin}}, \bibinfo {author}
  {\bibfnamefont {M.}~\bibnamefont {Krenn}}, \bibinfo {author} {\bibfnamefont
  {L.~S.}\ \bibnamefont {Madsen}}, \bibinfo {author} {\bibfnamefont
  {M.}~\bibnamefont {M{\aa}nsson}}, \bibinfo {author} {\bibfnamefont {N.~P.}\
  \bibnamefont {Mauranyapin}}, \bibinfo {author} {\bibfnamefont {A.~I.}\
  \bibnamefont {Melvin}}, \bibinfo {author} {\bibfnamefont {E.~M.}\
  \bibnamefont {Rasel}}, \bibinfo {author} {\bibfnamefont {L.~E.}\ \bibnamefont
  {Reichl}}, \bibinfo {author} {\bibfnamefont {R.}~\bibnamefont {Yampolskiy}},
  \bibinfo {author} {\bibfnamefont {P.~B.}\ \bibnamefont {Yasskin}}, \bibinfo
  {author} {\bibfnamefont {A.}~\bibnamefont {Zeilinger}}, \ and\ \bibinfo
  {author} {\bibfnamefont {S.}~\bibnamefont {Lidstr{\"{o}}m}},\ }\href
  {\doibase 10.1088/1402-4896/ab7a35} {\bibfield  {journal} {\bibinfo
  {journal} {Physica Scripta}\ }\textbf {\bibinfo {volume} {95}},\ \bibinfo
  {pages} {062501} (\bibinfo {year} {2020})}\BibitemShut {NoStop}%
\bibitem [{\citenamefont {{Rami Reddy}}\ \emph {et~al.}(2015)\citenamefont
  {{Rami Reddy}}, \citenamefont {Ravikumar}, \citenamefont {Nithya},\ and\
  \citenamefont {Gopukumar}}]{RamiReddy2015}%
  \BibitemOpen
  \bibfield  {author} {\bibinfo {author} {\bibfnamefont {B.~V.}\ \bibnamefont
  {{Rami Reddy}}}, \bibinfo {author} {\bibfnamefont {R.}~\bibnamefont
  {Ravikumar}}, \bibinfo {author} {\bibfnamefont {C.}~\bibnamefont {Nithya}}, \
  and\ \bibinfo {author} {\bibfnamefont {S.}~\bibnamefont {Gopukumar}},\ }\href
  {\doibase 10.1039/c5ta03173g} {\bibfield  {journal} {\bibinfo  {journal}
  {Journal of Materials Chemistry A}\ }\textbf {\bibinfo {volume} {3}},\
  \bibinfo {pages} {18059} (\bibinfo {year} {2015})}\BibitemShut {NoStop}%
\bibitem [{\citenamefont {Rai}\ \emph {et~al.}(2014)\citenamefont {Rai},
  \citenamefont {Anh}, \citenamefont {Gim}, \citenamefont {Mathew},\ and\
  \citenamefont {Kim}}]{Rai2014}%
  \BibitemOpen
  \bibfield  {author} {\bibinfo {author} {\bibfnamefont {A.~K.}\ \bibnamefont
  {Rai}}, \bibinfo {author} {\bibfnamefont {L.~T.}\ \bibnamefont {Anh}},
  \bibinfo {author} {\bibfnamefont {J.}~\bibnamefont {Gim}}, \bibinfo {author}
  {\bibfnamefont {V.}~\bibnamefont {Mathew}}, \ and\ \bibinfo {author}
  {\bibfnamefont {J.}~\bibnamefont {Kim}},\ }\href {\doibase
  10.1016/j.ceramint.2013.08.013} {\bibfield  {journal} {\bibinfo  {journal}
  {Ceramics International}\ }\textbf {\bibinfo {volume} {40}},\ \bibinfo
  {pages} {2411} (\bibinfo {year} {2014})}\BibitemShut {NoStop}%
\bibitem [{\citenamefont {Gupta}\ \emph {et~al.}(2013)\citenamefont {Gupta},
  \citenamefont {{Buddie Mullins}},\ and\ \citenamefont
  {Goodenough}}]{Gupta2013a}%
  \BibitemOpen
  \bibfield  {author} {\bibinfo {author} {\bibfnamefont {A.}~\bibnamefont
  {Gupta}}, \bibinfo {author} {\bibfnamefont {C.}~\bibnamefont {{Buddie
  Mullins}}}, \ and\ \bibinfo {author} {\bibfnamefont {J.~B.}\ \bibnamefont
  {Goodenough}},\ }\href {\doibase 10.1016/j.jpowsour.2013.06.073} {\bibfield
  {journal} {\bibinfo  {journal} {Journal of Power Sources}\ }\textbf {\bibinfo
  {volume} {243}},\ \bibinfo {pages} {817} (\bibinfo {year}
  {2013})}\BibitemShut {NoStop}%
\bibitem [{\citenamefont {Berthelot}\ \emph {et~al.}(2012)\citenamefont
  {Berthelot}, \citenamefont {Schmidt}, \citenamefont {Sleight},\ and\
  \citenamefont {Subramanian}}]{Berthelot2012}%
  \BibitemOpen
  \bibfield  {author} {\bibinfo {author} {\bibfnamefont {R.}~\bibnamefont
  {Berthelot}}, \bibinfo {author} {\bibfnamefont {W.}~\bibnamefont {Schmidt}},
  \bibinfo {author} {\bibfnamefont {A.~W.}\ \bibnamefont {Sleight}}, \ and\
  \bibinfo {author} {\bibfnamefont {M.~A.}\ \bibnamefont {Subramanian}},\
  }\href {\doibase 10.1111/cod.12358} {\bibfield  {journal} {\bibinfo
  {journal} {Journal of Solid State Chemistry}\ }\textbf {\bibinfo {volume}
  {196}},\ \bibinfo {pages} {225} (\bibinfo {year} {2012})}\BibitemShut
  {NoStop}%
\bibitem [{\citenamefont {Schulze}\ \emph {et~al.}(2008)\citenamefont
  {Schulze}, \citenamefont {H{\"{a}}fliger}, \citenamefont {Niedermayer},
  \citenamefont {Mattenberger}, \citenamefont {Bubenhofer},\ and\ \citenamefont
  {Batlogg}}]{Schulze2008}%
  \BibitemOpen
  \bibfield  {author} {\bibinfo {author} {\bibfnamefont {T.~F.}\ \bibnamefont
  {Schulze}}, \bibinfo {author} {\bibfnamefont {P.~S.}\ \bibnamefont
  {H{\"{a}}fliger}}, \bibinfo {author} {\bibfnamefont {C.}~\bibnamefont
  {Niedermayer}}, \bibinfo {author} {\bibfnamefont {K.}~\bibnamefont
  {Mattenberger}}, \bibinfo {author} {\bibfnamefont {S.}~\bibnamefont
  {Bubenhofer}}, \ and\ \bibinfo {author} {\bibfnamefont {B.}~\bibnamefont
  {Batlogg}},\ }\href {\doibase 10.1103/PhysRevLett.100.026407} {\bibfield
  {journal} {\bibinfo  {journal} {Physical Review Letters}\ }\textbf {\bibinfo
  {volume} {100}},\ \bibinfo {pages} {026407} (\bibinfo {year}
  {2008})}\BibitemShut {NoStop}%
\bibitem [{\citenamefont {Sugiyama}\ \emph {et~al.}(2011)\citenamefont
  {Sugiyama}, \citenamefont {Nozaki}, \citenamefont {Harada}, \citenamefont
  {Kamazawa}, \citenamefont {Ofer}, \citenamefont {M{\aa}nsson}, \citenamefont
  {Brewer}, \citenamefont {Ansaldo}, \citenamefont {Chow}, \citenamefont
  {Ikedo}, \citenamefont {Miyake}, \citenamefont {Ohishi}, \citenamefont
  {Watanabe}, \citenamefont {Kobayashi},\ and\ \citenamefont
  {Kanno}}]{Sugiyama2011}%
  \BibitemOpen
  \bibfield  {author} {\bibinfo {author} {\bibfnamefont {J.}~\bibnamefont
  {Sugiyama}}, \bibinfo {author} {\bibfnamefont {H.}~\bibnamefont {Nozaki}},
  \bibinfo {author} {\bibfnamefont {M.}~\bibnamefont {Harada}}, \bibinfo
  {author} {\bibfnamefont {K.}~\bibnamefont {Kamazawa}}, \bibinfo {author}
  {\bibfnamefont {O.}~\bibnamefont {Ofer}}, \bibinfo {author} {\bibfnamefont
  {M.}~\bibnamefont {M{\aa}nsson}}, \bibinfo {author} {\bibfnamefont {J.~H.}\
  \bibnamefont {Brewer}}, \bibinfo {author} {\bibfnamefont {E.~J.}\
  \bibnamefont {Ansaldo}}, \bibinfo {author} {\bibfnamefont {K.~H.}\
  \bibnamefont {Chow}}, \bibinfo {author} {\bibfnamefont {Y.}~\bibnamefont
  {Ikedo}}, \bibinfo {author} {\bibfnamefont {Y.}~\bibnamefont {Miyake}},
  \bibinfo {author} {\bibfnamefont {K.}~\bibnamefont {Ohishi}}, \bibinfo
  {author} {\bibfnamefont {I.}~\bibnamefont {Watanabe}}, \bibinfo {author}
  {\bibfnamefont {G.}~\bibnamefont {Kobayashi}}, \ and\ \bibinfo {author}
  {\bibfnamefont {R.}~\bibnamefont {Kanno}},\ }\href {\doibase
  10.1103/PhysRevB.84.054430} {\bibfield  {journal} {\bibinfo  {journal}
  {Physical Review B}\ }\textbf {\bibinfo {volume} {84}},\ \bibinfo {pages}
  {054430} (\bibinfo {year} {2011})}\BibitemShut {NoStop}%
\bibitem [{\citenamefont {Sugiyama}\ \emph
  {et~al.}(2009{\natexlab{a}})\citenamefont {Sugiyama}, \citenamefont
  {M{\aa}nsson}, \citenamefont {Ikedo}, \citenamefont {Goko}, \citenamefont
  {Mukai}, \citenamefont {Andreica}, \citenamefont {Amato}, \citenamefont
  {Ariyoshi},\ and\ \citenamefont {Ohzuku}}]{LiCrO2}%
  \BibitemOpen
  \bibfield  {author} {\bibinfo {author} {\bibfnamefont {J.}~\bibnamefont
  {Sugiyama}}, \bibinfo {author} {\bibfnamefont {M.}~\bibnamefont
  {M{\aa}nsson}}, \bibinfo {author} {\bibfnamefont {Y.}~\bibnamefont {Ikedo}},
  \bibinfo {author} {\bibfnamefont {T.}~\bibnamefont {Goko}}, \bibinfo {author}
  {\bibfnamefont {K.}~\bibnamefont {Mukai}}, \bibinfo {author} {\bibfnamefont
  {D.}~\bibnamefont {Andreica}}, \bibinfo {author} {\bibfnamefont
  {A.}~\bibnamefont {Amato}}, \bibinfo {author} {\bibfnamefont
  {K.}~\bibnamefont {Ariyoshi}}, \ and\ \bibinfo {author} {\bibfnamefont
  {T.}~\bibnamefont {Ohzuku}},\ }\href {\doibase 10.1103/PhysRevB.79.184411}
  {\bibfield  {journal} {\bibinfo  {journal} {Physical Review B}\ }\textbf
  {\bibinfo {volume} {79}},\ \bibinfo {pages} {184411} (\bibinfo {year}
  {2009}{\natexlab{a}})}\BibitemShut {NoStop}%
\bibitem [{\citenamefont {Sugiyama}\ \emph
  {et~al.}(2009{\natexlab{b}})\citenamefont {Sugiyama}, \citenamefont {Mukai},
  \citenamefont {Ikedo}, \citenamefont {Nozaki}, \citenamefont {M{\aa}nsson},\
  and\ \citenamefont {Watanabe}}]{Sugiyama2009}%
  \BibitemOpen
  \bibfield  {author} {\bibinfo {author} {\bibfnamefont {J.}~\bibnamefont
  {Sugiyama}}, \bibinfo {author} {\bibfnamefont {K.}~\bibnamefont {Mukai}},
  \bibinfo {author} {\bibfnamefont {Y.}~\bibnamefont {Ikedo}}, \bibinfo
  {author} {\bibfnamefont {H.}~\bibnamefont {Nozaki}}, \bibinfo {author}
  {\bibfnamefont {M.}~\bibnamefont {M{\aa}nsson}}, \ and\ \bibinfo {author}
  {\bibfnamefont {I.}~\bibnamefont {Watanabe}},\ }\href {\doibase
  10.1103/PhysRevLett.103.147601} {\bibfield  {journal} {\bibinfo  {journal}
  {Physical Review Letters}\ }\textbf {\bibinfo {volume} {103}},\ \bibinfo
  {pages} {147601} (\bibinfo {year} {2009}{\natexlab{b}})},\ \Eprint
  {http://arxiv.org/abs/0909.2921} {arXiv:0909.2921} \BibitemShut {NoStop}%
\bibitem [{\citenamefont {Sugiyama}\ \emph {et~al.}(2012)\citenamefont
  {Sugiyama}, \citenamefont {Nozaki}, \citenamefont {Harada}, \citenamefont
  {Kamazawa}, \citenamefont {Ikedo}, \citenamefont {Miyake}, \citenamefont
  {Ofer}, \citenamefont {M{\aa}nsson}, \citenamefont {Ansaldo}, \citenamefont
  {Chow}, \citenamefont {Kobayashi},\ and\ \citenamefont
  {Kanno}}]{Sugiyama2012}%
  \BibitemOpen
  \bibfield  {author} {\bibinfo {author} {\bibfnamefont {J.}~\bibnamefont
  {Sugiyama}}, \bibinfo {author} {\bibfnamefont {H.}~\bibnamefont {Nozaki}},
  \bibinfo {author} {\bibfnamefont {M.}~\bibnamefont {Harada}}, \bibinfo
  {author} {\bibfnamefont {K.}~\bibnamefont {Kamazawa}}, \bibinfo {author}
  {\bibfnamefont {Y.}~\bibnamefont {Ikedo}}, \bibinfo {author} {\bibfnamefont
  {Y.}~\bibnamefont {Miyake}}, \bibinfo {author} {\bibfnamefont
  {O.}~\bibnamefont {Ofer}}, \bibinfo {author} {\bibfnamefont {M.}~\bibnamefont
  {M{\aa}nsson}}, \bibinfo {author} {\bibfnamefont {E.~J.}\ \bibnamefont
  {Ansaldo}}, \bibinfo {author} {\bibfnamefont {K.~H.}\ \bibnamefont {Chow}},
  \bibinfo {author} {\bibfnamefont {G.}~\bibnamefont {Kobayashi}}, \ and\
  \bibinfo {author} {\bibfnamefont {R.}~\bibnamefont {Kanno}},\ }\href
  {\doibase 10.1103/PhysRevB.85.054111} {\bibfield  {journal} {\bibinfo
  {journal} {Physical Review B}\ }\textbf {\bibinfo {volume} {85}},\ \bibinfo
  {pages} {054111} (\bibinfo {year} {2012})}\BibitemShut {NoStop}%
\bibitem [{\citenamefont {M{\aa}nsson}\ and\ \citenamefont
  {Sugiyama}(2013)}]{Mansson2013}%
  \BibitemOpen
  \bibfield  {author} {\bibinfo {author} {\bibfnamefont {M.}~\bibnamefont
  {M{\aa}nsson}}\ and\ \bibinfo {author} {\bibfnamefont {J.}~\bibnamefont
  {Sugiyama}},\ }\href@noop {} {\bibfield  {journal} {\bibinfo  {journal}
  {Physica Scripta}\ }\textbf {\bibinfo {volume} {88}},\ \bibinfo {pages}
  {068509} (\bibinfo {year} {2013})}\BibitemShut {NoStop}%
\bibitem [{\citenamefont {Umegaki}\ \emph {et~al.}(2018)\citenamefont
  {Umegaki}, \citenamefont {Nozaki}, \citenamefont {Harada}, \citenamefont
  {M{\aa}nsson}, \citenamefont {Sakurai}, \citenamefont {Kawasaki},
  \citenamefont {Watanabe},\ and\ \citenamefont {Sugiyama}}]{Umegaki2018}%
  \BibitemOpen
  \bibfield  {author} {\bibinfo {author} {\bibfnamefont {I.}~\bibnamefont
  {Umegaki}}, \bibinfo {author} {\bibfnamefont {H.}~\bibnamefont {Nozaki}},
  \bibinfo {author} {\bibfnamefont {M.}~\bibnamefont {Harada}}, \bibinfo
  {author} {\bibfnamefont {M.}~\bibnamefont {M{\aa}nsson}}, \bibinfo {author}
  {\bibfnamefont {H.}~\bibnamefont {Sakurai}}, \bibinfo {author} {\bibfnamefont
  {I.}~\bibnamefont {Kawasaki}}, \bibinfo {author} {\bibfnamefont
  {I.}~\bibnamefont {Watanabe}}, \ and\ \bibinfo {author} {\bibfnamefont
  {J.}~\bibnamefont {Sugiyama}},\ }\href {\doibase 10.7566/jpscp.21.011018}
  {\bibfield  {journal} {\bibinfo  {journal} {JPS Conference Proceeding}\
  }\textbf {\bibinfo {volume} {21}},\ \bibinfo {pages} {011018} (\bibinfo
  {year} {2018})}\BibitemShut {NoStop}%
\bibitem [{\citenamefont {Alloul}\ \emph {et~al.}(2012)\citenamefont {Alloul},
  \citenamefont {Mukhamedshin}, \citenamefont {Dooglav}, \citenamefont
  {Dmitriev}, \citenamefont {Ciomaga}, \citenamefont {Pinsard-Gaudart},\ and\
  \citenamefont {Collin}}]{Alloul2012}%
  \BibitemOpen
  \bibfield  {author} {\bibinfo {author} {\bibfnamefont {H.}~\bibnamefont
  {Alloul}}, \bibinfo {author} {\bibfnamefont {I.~R.}\ \bibnamefont
  {Mukhamedshin}}, \bibinfo {author} {\bibfnamefont {A.~V.}\ \bibnamefont
  {Dooglav}}, \bibinfo {author} {\bibfnamefont {Y.~V.}\ \bibnamefont
  {Dmitriev}}, \bibinfo {author} {\bibfnamefont {V.~C.}\ \bibnamefont
  {Ciomaga}}, \bibinfo {author} {\bibfnamefont {L.}~\bibnamefont
  {Pinsard-Gaudart}}, \ and\ \bibinfo {author} {\bibfnamefont {G.}~\bibnamefont
  {Collin}},\ }\href {\doibase 10.1103/PhysRevB.85.134433} {\bibfield
  {journal} {\bibinfo  {journal} {Physical Review B}\ }\textbf {\bibinfo
  {volume} {85}},\ \bibinfo {pages} {134433} (\bibinfo {year}
  {2012})}\BibitemShut {NoStop}%
\bibitem [{\citenamefont {Siegel}\ \emph {et~al.}(2001)\citenamefont {Siegel},
  \citenamefont {Hirschinger}, \citenamefont {Carlier}, \citenamefont {Matar},
  \citenamefont {M{\'{e}}n{\'{e}}trier},\ and\ \citenamefont
  {Delmas}}]{Siegel2001}%
  \BibitemOpen
  \bibfield  {author} {\bibinfo {author} {\bibfnamefont {R.}~\bibnamefont
  {Siegel}}, \bibinfo {author} {\bibfnamefont {J.}~\bibnamefont {Hirschinger}},
  \bibinfo {author} {\bibfnamefont {D.}~\bibnamefont {Carlier}}, \bibinfo
  {author} {\bibfnamefont {S.}~\bibnamefont {Matar}}, \bibinfo {author}
  {\bibfnamefont {M.}~\bibnamefont {M{\'{e}}n{\'{e}}trier}}, \ and\ \bibinfo
  {author} {\bibfnamefont {C.}~\bibnamefont {Delmas}},\ }\href {\doibase
  10.1021/jp003832s} {\bibfield  {journal} {\bibinfo  {journal} {Journal of
  Physical Chemistry B}\ }\textbf {\bibinfo {volume} {105}},\ \bibinfo {pages}
  {4166} (\bibinfo {year} {2001})}\BibitemShut {NoStop}%
\bibitem [{\citenamefont {Masese}\ and\ \citenamefont {Al}(2020)}]{Titus2020}%
  \BibitemOpen
  \bibfield  {author} {\bibinfo {author} {\bibfnamefont {T.}~\bibnamefont
  {Masese}}\ and\ \bibinfo {author} {\bibfnamefont {E.}~\bibnamefont {Al}},\
  }\href@noop {} {\bibfield  {journal} {\bibinfo  {journal} {private
  communication}\ } (\bibinfo {year} {2020})}\BibitemShut {NoStop}%
\bibitem [{\citenamefont {Medarde}\ \emph {et~al.}(2013)\citenamefont
  {Medarde}, \citenamefont {Mena}, \citenamefont {Gavilano}, \citenamefont
  {Pomjakushina}, \citenamefont {Sugiyama}, \citenamefont {Kamazawa},
  \citenamefont {Pomjakushin}, \citenamefont {Sheptyakov}, \citenamefont
  {Batlogg}, \citenamefont {Ott}, \citenamefont {M{\aa}nsson},\ and\
  \citenamefont {Juranyi}}]{Medarde2013}%
  \BibitemOpen
  \bibfield  {author} {\bibinfo {author} {\bibfnamefont {M.}~\bibnamefont
  {Medarde}}, \bibinfo {author} {\bibfnamefont {M.}~\bibnamefont {Mena}},
  \bibinfo {author} {\bibfnamefont {J.~L.}\ \bibnamefont {Gavilano}}, \bibinfo
  {author} {\bibfnamefont {E.}~\bibnamefont {Pomjakushina}}, \bibinfo {author}
  {\bibfnamefont {J.}~\bibnamefont {Sugiyama}}, \bibinfo {author}
  {\bibfnamefont {K.}~\bibnamefont {Kamazawa}}, \bibinfo {author}
  {\bibfnamefont {V.~Y.}\ \bibnamefont {Pomjakushin}}, \bibinfo {author}
  {\bibfnamefont {D.}~\bibnamefont {Sheptyakov}}, \bibinfo {author}
  {\bibfnamefont {B.}~\bibnamefont {Batlogg}}, \bibinfo {author} {\bibfnamefont
  {H.~R.}\ \bibnamefont {Ott}}, \bibinfo {author} {\bibfnamefont
  {M.}~\bibnamefont {M{\aa}nsson}}, \ and\ \bibinfo {author} {\bibfnamefont
  {F.}~\bibnamefont {Juranyi}},\ }\href {\doibase
  10.1103/PhysRevLett.110.266401} {\bibfield  {journal} {\bibinfo  {journal}
  {Physical Review Letters}\ }\textbf {\bibinfo {volume} {110}},\ \bibinfo
  {pages} {266401} (\bibinfo {year} {2013})},\ \Eprint
  {http://arxiv.org/abs/1301.5827} {arXiv:1301.5827} \BibitemShut {NoStop}%
\bibitem [{\citenamefont {Benedek}\ \emph {et~al.}(2019)\citenamefont
  {Benedek}, \citenamefont {Yazdani}, \citenamefont {Chen}, \citenamefont
  {Wenzler}, \citenamefont {Juranyi}, \citenamefont {M{\aa}nsson},
  \citenamefont {Islam},\ and\ \citenamefont {Wood}}]{Benedek2019}%
  \BibitemOpen
  \bibfield  {author} {\bibinfo {author} {\bibfnamefont {P.}~\bibnamefont
  {Benedek}}, \bibinfo {author} {\bibfnamefont {N.}~\bibnamefont {Yazdani}},
  \bibinfo {author} {\bibfnamefont {H.}~\bibnamefont {Chen}}, \bibinfo {author}
  {\bibfnamefont {N.}~\bibnamefont {Wenzler}}, \bibinfo {author} {\bibfnamefont
  {F.}~\bibnamefont {Juranyi}}, \bibinfo {author} {\bibfnamefont
  {M.}~\bibnamefont {M{\aa}nsson}}, \bibinfo {author} {\bibfnamefont {M.~S.}\
  \bibnamefont {Islam}}, \ and\ \bibinfo {author} {\bibfnamefont {V.~C.}\
  \bibnamefont {Wood}},\ }\href {\doibase 10.1039/c8se00389k} {\bibfield
  {journal} {\bibinfo  {journal} {Sustainable Energy and Fuels}\ }\textbf
  {\bibinfo {volume} {3}},\ \bibinfo {pages} {508} (\bibinfo {year}
  {2019})}\BibitemShut {NoStop}%
\bibitem [{\citenamefont {Benedek}\ \emph {et~al.}(2020)\citenamefont
  {Benedek}, \citenamefont {Forslund}, \citenamefont {Nocerino}, \citenamefont
  {Yazdani}, \citenamefont {Matsubara}, \citenamefont {Sassa}, \citenamefont
  {Jur{\`{a}}nyi}, \citenamefont {Medarde}, \citenamefont {Telling},
  \citenamefont {M{\aa}nsson},\ and\ \citenamefont {Wood}}]{Benedek2020}%
  \BibitemOpen
  \bibfield  {author} {\bibinfo {author} {\bibfnamefont {P.}~\bibnamefont
  {Benedek}}, \bibinfo {author} {\bibfnamefont {O.~K.}\ \bibnamefont
  {Forslund}}, \bibinfo {author} {\bibfnamefont {E.}~\bibnamefont {Nocerino}},
  \bibinfo {author} {\bibfnamefont {N.}~\bibnamefont {Yazdani}}, \bibinfo
  {author} {\bibfnamefont {N.}~\bibnamefont {Matsubara}}, \bibinfo {author}
  {\bibfnamefont {Y.}~\bibnamefont {Sassa}}, \bibinfo {author} {\bibfnamefont
  {F.}~\bibnamefont {Jur{\`{a}}nyi}}, \bibinfo {author} {\bibfnamefont
  {M.}~\bibnamefont {Medarde}}, \bibinfo {author} {\bibfnamefont
  {M.}~\bibnamefont {Telling}}, \bibinfo {author} {\bibfnamefont
  {M.}~\bibnamefont {M{\aa}nsson}}, \ and\ \bibinfo {author} {\bibfnamefont
  {V.}~\bibnamefont {Wood}},\ }\href {\doibase 10.1021/acsami.9b21470}
  {\bibfield  {journal} {\bibinfo  {journal} {ACS Applied Materials and
  Interfaces}\ }\textbf {\bibinfo {volume} {12}},\ \bibinfo {pages} {16243}
  (\bibinfo {year} {2020})}\BibitemShut {NoStop}%
\bibitem [{\citenamefont {Yonemura}\ \emph {et~al.}(2014)\citenamefont
  {Yonemura}, \citenamefont {Mori}, \citenamefont {Kamiyama}, \citenamefont
  {Fukunaga}, \citenamefont {Torii}, \citenamefont {Nagao}, \citenamefont
  {Ishikawa}, \citenamefont {Onodera}, \citenamefont {Adipranoto},
  \citenamefont {Arai}, \citenamefont {Uchimoto},\ and\ \citenamefont
  {Ogumi}}]{Yonemura2014}%
  \BibitemOpen
  \bibfield  {author} {\bibinfo {author} {\bibfnamefont {M.}~\bibnamefont
  {Yonemura}}, \bibinfo {author} {\bibfnamefont {K.}~\bibnamefont {Mori}},
  \bibinfo {author} {\bibfnamefont {T.}~\bibnamefont {Kamiyama}}, \bibinfo
  {author} {\bibfnamefont {T.}~\bibnamefont {Fukunaga}}, \bibinfo {author}
  {\bibfnamefont {S.}~\bibnamefont {Torii}}, \bibinfo {author} {\bibfnamefont
  {M.}~\bibnamefont {Nagao}}, \bibinfo {author} {\bibfnamefont
  {Y.}~\bibnamefont {Ishikawa}}, \bibinfo {author} {\bibfnamefont
  {Y.}~\bibnamefont {Onodera}}, \bibinfo {author} {\bibfnamefont {D.~S.}\
  \bibnamefont {Adipranoto}}, \bibinfo {author} {\bibfnamefont
  {H.}~\bibnamefont {Arai}}, \bibinfo {author} {\bibfnamefont {Y.}~\bibnamefont
  {Uchimoto}}, \ and\ \bibinfo {author} {\bibfnamefont {Z.}~\bibnamefont
  {Ogumi}},\ }\href@noop {} {\bibfield  {journal} {\bibinfo  {journal} {Journal
  of Physics: Conference Series}\ }\textbf {\bibinfo {volume} {502}},\ \bibinfo
  {pages} {012053} (\bibinfo {year} {2014})}\BibitemShut {NoStop}%
\bibitem [{\citenamefont
  {Rodr{\'{i}}guez-Carvajal}(1993)}]{Rodriguez-Carvajal1993a}%
  \BibitemOpen
  \bibfield  {author} {\bibinfo {author} {\bibfnamefont {J.}~\bibnamefont
  {Rodr{\'{i}}guez-Carvajal}},\ }\href {\doibase 10.1016/0921-4526(93)90108-I}
  {\bibfield  {journal} {\bibinfo  {journal} {Physica B}\ }\textbf {\bibinfo
  {volume} {192}},\ \bibinfo {pages} {55} (\bibinfo {year} {1993})}\BibitemShut
  {NoStop}%
\bibitem [{\citenamefont {Yaouanc}\ and\ \citenamefont {{Dalmas De
  R{\'{e}}otier}}(2011)}]{Yaouanc2011}%
  \BibitemOpen
  \bibfield  {author} {\bibinfo {author} {\bibfnamefont {A.}~\bibnamefont
  {Yaouanc}}\ and\ \bibinfo {author} {\bibfnamefont {P.}~\bibnamefont {{Dalmas
  De R{\'{e}}otier}}},\ }\href@noop {} {\emph {\bibinfo {title} {{Muon Spin
  Rotation, Relaxation, and Resonance Applications to Condensed Matter}}}}\
  (\bibinfo  {publisher} {Oxford University Press},\ \bibinfo {year}
  {2011})\BibitemShut {NoStop}%
\bibitem [{\citenamefont {Suter}\ and\ \citenamefont
  {Wojek}(2012)}]{Suter2012}%
  \BibitemOpen
  \bibfield  {author} {\bibinfo {author} {\bibfnamefont {A.}~\bibnamefont
  {Suter}}\ and\ \bibinfo {author} {\bibfnamefont {B.~M.}\ \bibnamefont
  {Wojek}},\ }\href {\doibase 10.1016/j.phpro.2012.04.042} {\bibfield
  {journal} {\bibinfo  {journal} {Physics Procedia}\ }\textbf {\bibinfo
  {volume} {30}},\ \bibinfo {pages} {69} (\bibinfo {year} {2012})},\ \Eprint
  {http://arxiv.org/abs/1111.1569} {arXiv:1111.1569} \BibitemShut {NoStop}%
\end{thebibliography}%

\section*{Acknowledgements}
The authors thank P.~Gratrex (KTH Royal Institute of Technology) for his great support during the \musr~experiment. This research was supported by the Swedish Research Council (VR) through a Neutron Project Grant (Dnr. 2016-06955) as well as the Carl Tryggers Foundation for Scientific Research (CTS-18:272). J.S. acknowledge support from Japan Society for the Promotion Science (JSPS) KAKENHI Grant No. JP18H01863. Y.S. is funded by the Swedish Research Council (VR) through a Starting Grant (Dnr. 2017-05078). Y.S. and K.P acknowledge Chalmers Area of Advance-Materials science. E.N. is fully funded by the Swedish Foundation for Strategic Research (SSF) within the Swedish national graduate school in neutron scattering (SwedNess). D.A. acknowledges partial financial support from the Romanian UEFISCDI project PN-III-P4-ID-PCCF-2016-0112, Contract Nr. 6/2018. T.M. acknowledges the National Institute of Advanced Industrial Science Technology (AIST), Japan Society for the Promotion of Science(JSPS KAKENHI Grant Numbers 19 K15685) and Japan Prize Foundation. Finally, the authors are grateful to J-PARC, Paul Scherrer Institute and ISIS/RAL for the allocated muon/neutron beam-time as well as the great support from their technical staff.
\section*{Author contributions statement}

M.M. and T.M. conceived the experiments. N.M., E.N., O.K.F., A.Z., K.P., D.A., J.S., Y.S., and M.M. prepared and conducted all the experiments. Z.G., S.P.C., T.S., T.K. and A.K. supported the neutron, muon and magnetization experiments. N.M., O.K.F., D.A., E.N., R.P. and A.Z., analyzed the results. T.M. synthesized the samples and conducted the initial sample characterizations. N.M. and M.M. created the first draft, and all authors reviewed and contributed to the final manuscript in several steps.​

\section*{Additional information}
\textbf{Supporting Information} : Supporting Information is available. \\
\textbf{Competing interests} : The authors declare no competing interests.

\newpage

\newpage

Supporting Information for:\\
  \large{Magnetism and Ion Diffusion in Honeycomb Layered Oxide \knto: First Time Study by Muon Spin Rotation \& Neutron Scattering}

\begin{table}[h]
\caption{\label{tablestructure} Structural parameters of \knto~at 300 K from Reitveld refinements of XRPD and NPD data. The space group is $P6_3/mcm$. Te atoms are in the (2$b$) Wyckoff position (WP) [0 0 0], Ni in (4$d$) [2/3 1/3 0], O in (12$k$) [$x x z$], K1 in (6$g$)[x 0 1/4], K2 in (4$c$) [1/3 2/3 1/4] and K3 in (2$a$) [0 0 1/4]. The refined $x$ and $z$ parameters and isotropic thermal factors ($B$) are given for the different ions. The occupancy of K sites are also refined ($Oc$). Maximum values among the refinement results of different banks (global fit)\textit{R}$_{Bragg}$ is described here. } 

\begin{tabular}{ccdd}
\hline
&&
\multicolumn{1}{c}{\textrm{NPD}}\\
\hline
& \ac~(\AA) & 5.258 (1) \\
& \cc~(\AA) &  12.42 (1) \\
& \Vc~(\AA$^3$) & 297.3 (1) \\
Te& \textit{B} (\AA$^2$)  &  0.33 (10)\\
Ni& \textit{B} (\AA$^2$)  &    0.45 (3)\\
O& \x  & 0.6887 (3) \\
& $z$  & 0.5843 (1) \\
& \textit{B} (\AA$^2$)  & 0.47 (2)\\
K1& $x$  & 0.3657 (19) \\
& \textit{B} (\AA$^2$)  & 3.89 (33) \\
& \textit{Oc}  & 0.494 (1) \\
K2& \textit{B} (\AA$^2$)  &  3.89 (33) \\
& \textit{Oc}  &  0.272 (1) \\
K3& \textit{B} (\AA$^2$)  &  3.89 (33) \\
& \textit{Oc}  &  0.020(1) \\
& \textit{R}$_{Bragg}$ ($\%$)   & 6.56 \\
\hline
\end{tabular}

\end{table}

%

\begin{table}[h]
\caption{\label{tablelength} Selected distances (\AA) of \knto~obtained from XRPD and NPD data}
\begin{tabular}{ccdd}
\hline
&
\multicolumn{1}{c}{\textrm{NPD}}\\
\hline
Te - O        & 1.943 (2)     \\
Ni - O       & 2.094 (2)     \\
K1 - K2         & 1.674 (4)    \\
K1 - K3          & 1.923 (10)     \\
Ni - Ni          & 3.036 (1)     \\
\hline
\end{tabular}
\end{table}

 \begin{figure}[h]
   \begin{center}
     \includegraphics[width=75mm]{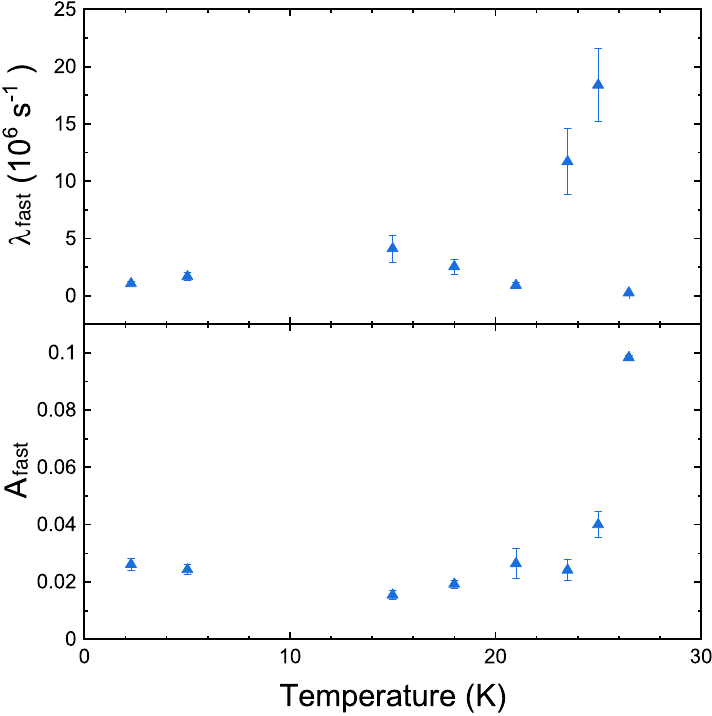}
   \end{center}
  \caption{Temperature dependencies of \musr~parameters taken for \knto; (top) the relaxation
rates ($\lambda_{fast}$) and (bottom) the asymmetries (A$_{\rm fast}$). The data were obtained by fitting the ZF spectra using Eq.2 described in the main text. 
}
   \label{ZF_para}
 \end{figure}

 \begin{figure}[h]
   \begin{center}
     \includegraphics[width=75mm]{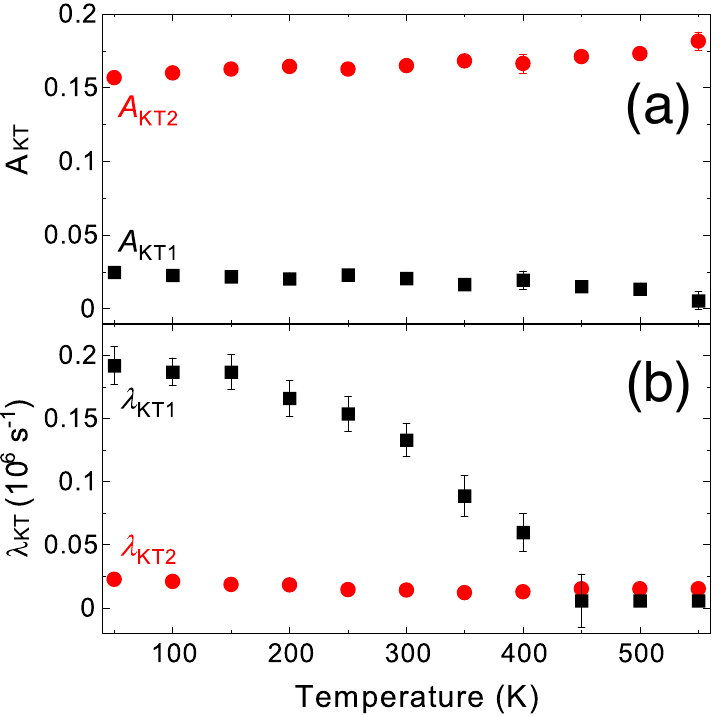}
   \end{center}
  \caption{The temperature dependencies of (a) the asymmetry ($A_{\rm KT}$) and (b) $\lambda_{\rm KT}$ for \knto.}
   \label{ZF_para}
 \end{figure}


\end{document}